\documentclass[journal=jctcce,manuscript=article]{achemso}
\usepackage{amsmath,amssymb}
\usepackage{graphicx}
\usepackage{booktabs}
\usepackage[version=4]{mhchem}
\usepackage{siunitx}
\usepackage{xurl}
\usepackage[hidelinks]{hyperref}
\graphicspath{{figures/261015/}}

\title{From Algebraic Completeness to Orbit Separation: Recursive Separating Invariants for Global Neural Network Potential Energy Surfaces}
\author{Yiping Hao}
\altaffiliation{These authors contributed equally to this work.}
\affiliation{State Key Laboratory of Chemical Reaction Dynamics, Dalian Institute of Chemical Physics, Chinese Academy of Science, Dalian 116023, People’s Republic of China}
\alsoaffiliation{School of Chemical Sciences, University of Chinese Academy of Sciences, Beijing 100049, China}
\author{Jing Lou}
\altaffiliation{These authors contributed equally to this work.}

\affiliation{State Key Laboratory of Chemical Reaction Dynamics, Dalian Institute of Chemical Physics, Chinese Academy of Science, Dalian 116023, People’s Republic of China}
\alsoaffiliation{School of Chemical Sciences, University of Chinese Academy of Sciences, Beijing 100049, China}
\author{Bina Fu}
\email{bina@dicp.ac.cn}
\affiliation{State Key Laboratory of Chemical Reaction Dynamics, Dalian Institute of Chemical Physics, Chinese Academy of Science, Dalian 116023, People’s Republic of China}
\alsoaffiliation{Hefei National Laboratory, Hefei 230088, China}
\alsoaffiliation{School of Chemical Sciences, University of Chinese Academy of Sciences, Beijing 100049, China}

\author{Dong H. Zhang}
\affiliation{State Key Laboratory of Chemical Reaction Dynamics, Dalian Institute of Chemical Physics, Chinese Academy of Science, Dalian 116023, People’s Republic of China}
\alsoaffiliation{Hefei National Laboratory, Hefei 230088, China}
\alsoaffiliation{School of Chemical Sciences, University of Chinese Academy of Sciences, Beijing 100049, China}

\begin{document}

\maketitle

\begin{abstract}
Permutationally invariant polynomials (PIPs) provide a systematically improvable, symmetry-preserving representation for global potential energy surfaces (PESs), but their high computational cost as neural network descriptors limits applications to large molecular systems. Fundamental invariants (FIs), which form an algebraically complete generating set, substantially reduce the number of descriptors and enable neural network PESs for larger systems. However, PES evaluation remains computationally expensive as molecular size and complexity increase. Here, we introduce recursive separating invariants (RSIs), shifting the descriptor design principle from algebraic completeness to orbit separation. RSIs are constructed systematically through correlation order and recursion depth, providing a more computationally efficient representation.Benchmarks on reactive and nonreactive systems demonstrate that RSI-based neural network (RSI-NN) PESs achieve approximately an order-of-magnitude faster evaluation than their FI-NN counterparts, with system-dependent prediction accuracy that is comparable in several cases. These results establish orbit separation as a promising principle for developing efficient symmetry-preserving descriptors for global neural network PESs.

\end{abstract}

\section{Introduction}

The potential energy surface (PES) is the central object connecting electronic structure theory with molecular dynamics simulation: its energy governs nuclear motion, while its gradient supplies the forces needed for dynamical propagation. Machine-learning potential energy surfaces (PESs) provide a route to molecular simulation at a fraction of the cost of direct electronic structure calculations.\cite{Unke2021MLFF} Their reliability and efficiency depend critically on predictive accuracy, computationally efficient evaluation, and the rigorous preservation of physical symmetries. Local atomic-environment models,\cite{Behler2007HDNNP} continuous-filter networks,\cite{Schutt2018SchNet} and equivariant message-passing models\cite{Batzner2022NequIP,Batatia2022MACE} have greatly expanded the range of accessible systems. Complementing these approaches, global molecular representations remain useful for gas-phase reactions, strongly coupled vibrations, and large-amplitude rearrangements, where a single descriptor encodes the complete molecular configuration.\cite{Qu2018PIPReview,Bowman2025PIPPerspective}

Permutationally invariant polynomials (PIPs) construct such representations by symmetrizing monomials of interatomic-distance variables under exchanges of identical atoms.\cite{Braams2009PIP,Xie2010MSA} When the underlying coordinates separate the relevant physical configurations, the Stone--Weierstrass theorem provides a systematic approximation route on a compact configuration domain\cite{Rudin1976Principles}. PIP neural networks (PIP-NNs) combine these symmetry-adapted inputs with nonlinear regression.\cite{Jiang2013PIPNN,Li2013PIPNN2} The cost of direct global expansions, however, grows rapidly: for $N$ atoms and $M=\binom{N}{2}$ pairwise variables, the number of candidate monomials through degree $d$ is $\binom{M+d}{d}$. This growth affects both basis construction and repeated evaluation.

%Global PIP representations and efficient monomial-symmetrization algorithms for constructing their bases were developed by Bowman and co-workers.\cite{Braams2009PIP,Xie2010MSA}. When the underlying coordinates separate the relevant physical configurations, the Stone--Weierstrass theorem\cite{Rudin1976Principles} provides a systematic route to approximating continuous permutation-invariant PESs on a compact configuration domain. PIP neural networks (PIP-NNs), which combine symmetry-adapted polynomial inputs with nonlinear neural-network regression, were introduced by Jiang and Guo.\cite{Jiang2013PIPNN} and Li et al. extended this approach to four-atom systems.\cite{Li2013PIPNN2} The cost of direct global expansions, however, grows rapidly: for $N$ atoms and $M=\binom{N}{2}$ pairwise variables, the number of candidate monomials through degree $d$ is $\binom{M+d}{d}$. This growth affects both basis construction and repeated evaluation.

Several developments address this computational burden. Automated generation and parallel implementations make increasingly complex expansions accessible,\cite{Moberg2021PIPUnrestricted} while basis compaction and reverse differentiation accelerate energy and gradient evaluation.\cite{Houston2022PIPGradients} Fragment-based constructions reduce the expansion problem through molecular decomposition,\cite{Conte2020EfficientPIP} and atomic PIPs organize polynomial representations around local environments.\cite{Allen2021aPIP} Fragmented PIP models have also been applied to the conformational space of long-chain alkanes.\cite{Qu2024C14H30} These advances motivate a complementary question: how can the evaluation cost be reduced while retaining a single global molecular descriptor?

Fundamental invariant neural networks (FI-NNs) approach this problem through invariant algebra, replacing a direct PIP basis with a finite algebraic generating set.\cite{Shao2016FINN} Subsequent work extended FI generation to systems containing up to ten atoms\cite{Chen2020FINN2} and established its use in reactive PES construction.\cite{Fu2023FIReview} More recent group-generator algorithms avoid explicit enumeration of every permutation,\cite{Hao2025PIPFIAlgorithms} addressing a major bottleneck in generation of invariant expressions. Efficient generation does not, by itself, remove the cost of evaluating the resulting expressions. This distinction suggests reconsidering the information a descriptor must retain, rather than requiring it to generate the entire invariant algebra.

Orbit separation provides such an information criterion: a descriptor distinguishes configurations that are inequivalent under the relevant symmetries. An algebraic generating set is necessarily separating, whereas a separating set need not generate the invariant algebra.\cite{Kemper2009Separating} Distance power sums illustrate both the opportunity and the limitation of this relaxation. They are inexpensive to evaluate recursively and, when complete, recover the unordered distance multiset. That multiset generically determines a point configuration under suitable conditions,\cite{Boutin2004Reconstructing,Boutin2007Distances} but it loses atomic-assignment information and cannot distinguish certain inequivalent configurations.

Here we introduce recursive separating invariants (RSIs), using orbit separation as a design objective for a computationally economical global representation. Low-correlation invariant templates couple a selected number of geometric variables, while univariate recursion extends their nonlinear order. Bond-angle channels supplement the joint geometric information retained by the radial descriptors. We develop the construction and its information content, then evaluate the radial-plus-bond-angle representation on seven nonreactive and five reactive benchmarks. The comparison uses FI-NNs with matched hidden-layer widths and assesses the best fitting errors obtained under a common protocol together with the time required to evaluate energies from Cartesian coordinates.

\section{Theory and Method}

The construction separates two design choices: which geometric variables are coupled in an invariant, and how their nonlinear dependence is expanded. Orbit separation motivates the choice of information to retain, while correlation order and recursion depth organize its evaluation. All finite truncations considered here preserve permutation invariance; global orbit separation is a design objective rather than a property assumed for every truncation.

\subsection{Algebraic Completeness and Orbit Separation}

Let $G$ be the permutation group of identical atoms acting on $M$ pairwise variables. A set of invariant polynomials $F=\{f_1,\ldots,f_{n_{\mathrm{feat}}}\}$ is algebraically complete if every $G$-invariant polynomial can be written as a polynomial in its elements. FIs form a minimal algebraic generating set. Thus, a function expressed through PIPs can also be expressed as a function of FIs, although the two representations need not be equally easy to fit with a finite neural network.

Orbit separation asks a weaker question. The descriptor is evaluated on the molecule's bond-length vector, $A=(r_{1,2},r_{1,3},\dots,r_{1,N},\dots,r_{N-1,N})$. The descriptor set $F$ defines the map
\[
\mathbf F(\vec{x})
=\bigl(f_1(\vec{x}),\ldots,f_{n_{\mathrm{feat}}}(\vec{x})\bigr)^{\mathsf T},
\]
and it is separating if and only if $\mathbf F(A)=\mathbf F(B)$ implies $A=\sigma B$ for some $\sigma\in G$. Invariant polynomials are constant on each orbit, so separation is equivalent to injectivity of $\mathbf F$ on the quotient space. Algebraic completeness implies this property, but the converse need not hold.\cite{Kemper2009Separating}

Formaldehyde provides a concrete example. The nontrivial permutation exchanges its two hydrogen atoms, and a minimal algebraic generating set is
\[
\begin{gathered}
F_{\mathrm{FI}}
=\{r_{\mathrm{CO}},r_{\mathrm{H_1H_2}},a,b,q,p,s\},\\
a=r_{\mathrm{CH_1}}+r_{\mathrm{CH_2}},\qquad
u=r_{\mathrm{CH_1}}-r_{\mathrm{CH_2}},\\
b=r_{\mathrm{OH_1}}+r_{\mathrm{OH_2}},\qquad
v=r_{\mathrm{OH_1}}-r_{\mathrm{OH_2}},\\
q=u^2-v^2,\qquad p=2uv,\qquad s=u^2+v^2.
\end{gathered}
\]
On the real domain, $s=\sqrt{q^2+p^2}$. Removing $s$ therefore leaves six invariants that still separate all real bond-length orbits, even though $s$ is not a polynomial in the remaining generators. Since separability is the weaker requirement, the minimal separating set may be smaller than the minimal algebraic generating set. SI Sections S1--S2 give the proof and the minimality arguments.

For PES fitting, the relevant domain is the set of physically realizable configurations, modulo the joint action of $G$ and the Euclidean group $E(3)$, including reflection. Separation on the full abstract distance space is sufficient but not necessary on this physical domain, because many combinations in the abstract bond-length space do not correspond to actual physical configurations. On a compact physical configuration domain, a continuous invariant descriptor that separates these orbits induces a continuous one-to-one map from the quotient space to its image, with a continuous inverse. Every continuous invariant potential on that domain can consequently be written as
\[
V=h\circ\mathbf F
\]
with continuous $h$ on the descriptor image. Separation is therefore sufficient to retain the information needed for arbitrary continuous invariant targets, without requiring polynomial generation of the full invariant algebra. This information criterion does not determine the approximation cost for a particular finite network.

\subsection{Power Sums: Recovery of the Bond-Value Multiset and Its Limitations}

Power sums are the simplest recursively evaluated invariant descriptors, and their information content can be characterized exactly. Taken over a scalar function of bond length, they are cheap to evaluate and provide a large amount of separability, though not strict separability. We use them as the seed of the construction.

For a function of $M$ variables, one can define the power sums

\[
p_n(x)=\sum_{i=1}^{M}x_i^n,
\qquad n=1,\ldots,M.
\]

If only the first $D$ orders are computed, recursion can be carried out variable by variable,

\[
x_i^{n+1}=x_i^n x_i,
\]

with accumulation at each order. The total time complexity is $O(MD)$, and the working memory can be kept at $O(M)$. When the recursion is carried out fully to $D=M$, the time complexity is $O(M^2)$.

For multi-element systems, all interatomic distances can be grouped according to element-pair category, and power sums constructed independently for each group. For simplicity of discussion, we assume below that the system contains only one element. In this case, $N$ identical atoms have $M=\binom N2$ distances, and the complete first $M$ power sums uniquely determine the unordered multiset of these bond values; a detailed proof is given in SI Section S3. By an unordered multiset we mean that only each bond value and its multiplicity are retained, without recording its position in the bond-length vector or which pair of atoms it belongs to; for example, the unordered multiset corresponding to $(a,b,a)$ is $\{\!\{a,a,b\}\!\}$. However, reducing a bond-length vector to an unordered multiset discards the positional and atomic-assignment information of each component, and therefore in general cannot guarantee global strict separability of the bond-length vector or of the physical configuration.

This loss of information can be understood as over-symmetrization. An arbitrary atomic permutation $S_N$ induces a permutation group $\rho(S_N)$ on the $M$ distances, but for $N\geq4$,

\[
\rho(S_N)\subsetneq S_M.
\]

The complete power sums are invariant under any rearrangement of all $M$ distances, and can therefore separate distinct orbits of $S_M$; however, the true molecular symmetry is only the smaller group $\rho(S_N)$. A single orbit of $S_M$ may contain multiple mutually inequivalent orbits of $\rho(S_N)$, which power sums cannot distinguish.

A second four-atom example shows what is lost when atomic-assignment information is removed. Consider four identical atoms and six pairwise distances. Take two positive numbers $a\neq b$, and define three edge-length assignments:

\[
\begin{aligned}
A:\quad &r_{12}=r_{13}=r_{14}=a,\\
B:\quad &r_{12}=r_{13}=r_{23}=a,\\
C:\quad &r_{12}=r_{23}=r_{34}=a,
\end{aligned}
\]

with the remaining three edges all equal to $b$. In these three assignments, the edges taking value $a$ form, respectively, a star, a triangle, and a three-edge path. All three have exactly the same unlabeled edge-value multiset

\[
\{a,a,a,b,b,b\},
\]

so for any $n\geq1$,

\[
p_n(A)=p_n(B)=p_n(C)=3a^n+3b^n.
\]

The three edge-length adjacency relations differ, and no atomic permutation maps one to another, so they lie in different $S_4$ configuration orbits. The counterexample is physical as well as combinatorial: with $a=\sqrt2,\ b=1$, all three assignments correspond to nondegenerate three-dimensional configurations. SI Section S5 gives their explicit realization and correlation quantities.

\subsection{Physical-Configuration Separability of Power Sums}

The four-atom counterexample of the previous section shows that even the complete first $M$ power sums cannot guarantee strict configurational separability over the entire bond-length vector space. Their separability is nevertheless stronger than the counterexample suggests, because physical distances are not an arbitrary set of positive numbers. A three-dimensional configuration of $N$ atoms typically has $3N-6$ degrees of freedom once overall translation and rotation are removed, whereas the number of pairwise distances is $M=\binom N2$. The difference between the two is

\[
M-(3N-6)=\frac{(N-3)(N-4)}2.
\]

When $N=4$, this difference is zero, and nondegenerate physical distances carry no additional equality-type constraints locally; this matches the counterexample of the previous section, where the same set of unlabeled distances can be assigned to the six edges of a tetrahedron in different ways and still yield a valid configuration. When $N\geq5$, the set of physical distances has more dimensions than the configuration itself, so the distances must satisfy additional geometric constraints, and most arbitrary rearrangements of bond values no longer correspond to three-dimensional configurations. For systems with more than four atoms, many of the additional equivalence relations that power sums introduce in the abstract bond-length space disappear once the domain is restricted to physical distances.

The configuration-reconstruction results of Boutin and Kemper, from invariant algebra, further show that for a three-dimensional configuration in general position with $N\geq5$, the complete unlabeled multiset of pairwise distances uniquely determines the physical configuration, up to overall translation, rotation, reflection, and relabeling of identical atoms. Here, ``general position'' means that the configuration does not lie in a low-dimensional algebraic exceptional set defined by special coordinate or distance relations; ambiguity in reconstruction from unlabeled distances can only arise within this exceptional set, which has Lebesgue measure zero in the continuous configuration space. A configuration distribution with a continuous probability density in the full Cartesian space therefore lands on this set with probability zero. SI Section S4 discusses the corresponding conditions for multi-element systems and for constrained sampling.

If the bond-length transformation $u(r)$ is injective, the complete first $M$ power sums can recover $\{\!\{u(r_{ij})\}\!\}$. Thus, the information contained in the power sums can be summarized as

\[
(p_1,\ldots,p_M)
\ \Longrightarrow\
\{\!\{u(r_{ij})\}\!\}
\ \Longrightarrow\
\text{physical configuration},
\qquad N\geq5,
\]

where the first arrow holds strictly, and the second arrow holds for configurations in general position. Hence, the complete first $M$ power sums separate physical configurations sampled randomly from a continuous distribution with probability $1$. The four-atom ambiguity therefore does not imply that complete distance power sums discard most physical configurational information.

However, probability-$1$ separability is not sufficient to guarantee that power sums form a stable and easily fittable potential energy descriptor. Denote the complete power-sum descriptor by $\Phi(Q)$. If two inequivalent configurations $A$ and $B$ satisfy

\[
\Phi(A)=\Phi(B)=\mathbf p,
\qquad
V(A)\neq V(B),
\]

then the potential energy cannot be defined as a single-valued function at the descriptor point $\mathbf p$: for configuration sequences $A_n$ and $B_n$ approaching $A$ and $B$, respectively, continuity of the descriptor gives

\[
\Phi(A_n),\Phi(B_n)\longrightarrow\mathbf p,
\]

while the energy can tend to two different limits. Consequently, within any small neighborhood of $\mathbf p$ there may appear configurations whose descriptors are very close but whose energies differ markedly. Even though the set of strict collisions has measure zero, it may still cause near-collisions and poorly conditioned inverse mappings in its neighborhood, thereby increasing the fitting difficulty for a finite-capacity regression model.

Thus, the complete power sums provide a representation that is strong yet still insufficient: they achieve generic physical separability at a relatively low correlation order, but strict separability and stability near collisions are still missing. The generic-separability result also requires the complete first $M$ power sums and cannot be directly extended to the finite-order truncation $D<M$ used in practice. These distinctions motivate adding geometric correlations rather than relying on recursion depth alone.

\subsection{Fundamental Correlation Templates and Recursive Enhancement}

For a monomial
\[
m=\prod_{e=1}^{M}x_e^{a_e},
\]
define its algebraic degree and correlation order as
\[
\deg(m)=\sum_{e=1}^{M}a_e,\qquad
\operatorname{corr}(m)=\left|\{e:a_e>0\}\right|.
\]
The correlation order counts distinct variables, not the number of atoms. Thus, $x_e^{100}$ has degree 100 but correlation order 1, whereas $x_{e_1}x_{e_2}x_{e_3}$ has degree and correlation order 3. Monomials in the same permutation orbit share both quantities.

A square-free monomial defines a fundamental correlation template
\[
B_{[S]}(x)=\sum_{S'\in G\cdot S}\prod_{e\in S'}x_e,
\qquad S=\{e_1,\ldots,e_k\},
\]
where each distinct subset in the orbit is counted once. Its degree and correlation order are both $k$; the term \textit{fundamental} here refers to the template, not to a minimal algebraic generating set.

The four-atom example illustrates why correlation order matters. Power sums cannot distinguish $A$, $B$, and $C$. Second-correlation-order orbit sums distinguish $C$ from $A$ and $B$, but no increase in their recursion depth separates the star from the triangle. A triangle-product orbit sum of correlation order 3 resolves that remaining ambiguity (SI Section S5). Recursion depth and correlation order therefore control different aspects of the representation.

The evaluation cost of an invariant, and the number of monomials it contains, track the correlation order rather than the algebraic degree: a high-degree term of low correlation order is usually cheaper. Power sums illustrate the point. One picks the cheapest first-correlation-order template, recurses on each variable to obtain $x_e^{a_e}$, and substitutes the results back into that template, which yields a whole family of power sums at low cost. To increase degree within a fixed template, first generate powers recursively and then substitute them into the orbit sum:
\[
x_e^{n+1}=x_ex_e^n,\qquad
B_n(\vec{x})=B(x_1^n,\ldots,x_M^n),\quad n=1,\ldots,D.
\]
Applying the same scalar transformation to exchangeable variables preserves the permutation symmetry of $B$. This construction reuses univariate intermediate values across the terms of an invariant and across templates, rather than evaluating every expanded high-degree monomial independently.

In multi-element systems, different element-pair categories may use different recursion orders, provided exchangeable variables are treated identically. Independent exponents on same-type factors require additional symmetrization: for example, $x_e^p x_f^q+x_e^q x_f^p$ remains unchanged under $e\leftrightarrow f$, whereas a single product with $p\ne q$ generally does not. Same-exponent recursion generates a subset of the higher-degree polynomials at a given correlation order. SI Section S6 specifies the orbit construction and its invariance.

Repeating the recursion with $k$-th-order templates produces large families of high-degree invariants; adding them to the descriptor set improves separability and reduces the PES fitting error. For same-exponent second-correlation-order terms, the $M=O(N^2)$ distances form at most $O(N^4)$ pairs. Direct recursion and summation through order $D$ therefore cost $O(N^4D)$ for a fixed choice of scalar-basis families, well below the cost of expanding a high-order PIP directly. The total cost also depends on the retained templates and basis combinations, as detailed in SI Section S9. This organization permits a diagonal truncation: conventional invariant fitting truncates by algebraic degree alone, whereas here the algebraic degree and the correlation order are truncated jointly, so that low-correlation templates can be assigned greater recursion depth while more expensive higher-correlation terms receive a smaller basis budget.

\subsection{Radial Functions and Bond-Angle Enhancement}

The recursion can start from scalar functions of distance, including reciprocal, exponential, and Gaussian-type mappings. Exchangeable distances use the same mapping and parameters, preserving the symmetry of the template. Several basis functions can also be used within a single template: each initializes the recursion separately, the results are substituted back into the same template, and the resulting descriptors are combined. Positive high-degree outputs can also be rescaled through root extraction or logarithmic transformations; SI Section S7 discusses their domains and numerical implications. The system-specific basis choices, recursion depths, and rescaling rules are specified in SI Section S10.

The construction is not restricted to bond lengths. Scalar functions of other internal coordinates can serve as the recursion variables, and once symmetrization over atomic permutations is imposed, a single monomial may involve several types of internal coordinates at once. Here we use first-correlation-order templates of bond angles and dihedral angles to build angular invariants that supplement the bond-length descriptors.

Bond angles provide a complementary way to encode correlations among distances. The cosine of $\angle ijk$ depends jointly on $r_{ij}$, $r_{jk}$, and $r_{ik}$, even when it enters only a first-correlation-order sum over angle variables. In the four-atom example, the sum of bond-angle cosines distinguishes the star from the triangle, and the first two Chebyshev angle sums distinguish all three configurations for $a=\sqrt{2}$ and $b=1$ (SI Section S5.4). This example connects the information lost by distance power sums to the angular channels used below.

Starting from $z=\cos\theta$, higher-order angular functions are generated by
\[
T_0(z)=1,\qquad T_1(z)=z,\qquad
T_{n+1}(z)=2zT_n(z)-T_{n-1}(z).
\]
Because $T_n(\cos\theta)=\cos(n\theta)$, neither explicit angle evaluation nor repeated trigonometric evaluation is required. Summing $T_n(z)$ over each orbit of equivalent bond angles preserves permutation invariance. The center and endpoint roles are retained when constructing these orbits. First correlation order here refers to the angle variables, not to the underlying distances or atomic body order.

The benchmark representation combines radial and bond-angle channels. Enumerating all bond angles and recursing through depth $D$ has a direct evaluation cost of $O(N^3D)$. Dihedral cosines admit an analogous construction at $O(N^4D)$, the same order as the upper bound for the same-exponent recursion over bond-length second-correlation-order terms, but are not included in the present benchmarks; SI Section S8 presents this extension and its geometric domain. SI Section S10 records the configurations used in the benchmarks, and Section S11 describes the radial-only versus radial-plus-bond-angle comparisons.

\section{Computational Details}

\subsection{Benchmark Systems}

We evaluate RSI using seven nonreactive and five reactive systems, with energy-fitting accuracy and potential energy evaluation time as the two principal performance measures. Throughout the benchmark comparison, RSI denotes the representation containing both radial and bond-angle channels. The radial and angular channel counts and system-specific recursion limits are listed in SI Tables~S3 and~S4, respectively; their construction and degree-reduction rules are described in SI Section S10.2.

The nonreactive benchmarks contain 9--21 atoms. Configurations were selected from molecular dynamics trajectories in the MD17 data family,\cite{Chmiela2017,Christensen2020} using a Euclidean-distance criterion applied to bond-length vectors, yielding approximately 50\,000--80\,000 configurations per system. The reactive benchmarks contain 9--15 atoms and comprise \ce{F + C2H6}, \ce{F^- + (CH3)3CI}, \ce{Cl^- + (CH3)3CI}, and the reactions of syn- and anti-\ce{CH3CHOO} with \ce{H2O}. The corresponding FI-based potential energy surfaces and dynamics studies were reported by Lu et al for \ce{F + C2H6}\cite{Lu2025Ethane} and the two tert-butyl iodide reactions,\cite{Lu2022Fluoride,Lu2026Chloride} and by Liu et al for the syn and anti Criegee intermediate reactions.\cite{Liu2025Syn,Liu2025Anti}

Table~\ref{tab:systems} summarizes the benchmark sizes and network architectures. The elemental ordering in the formula column follows that used to generate the FI descriptors. The \ce{Cl^- + (CH3)3CI} benchmark corresponds to the reaction-region data (Part 2) in the Supporting Information of the original study.\cite{Lu2026Chloride} A complete dataset here means all configurations used for the corresponding benchmark in Table~\ref{tab:systems}, including both its training and test subsets.

\begin{table}[htbp]
  \centering\small
  \caption{Benchmark systems and neural-network architectures}
  \label{tab:systems}
  \begin{tabular}{llrrl}
    \toprule
    System & Formula & Atoms & Configurations/$10^4$ & Network (FI/RSI) \\
    \midrule
    \multicolumn{5}{l}{Nonreactive benchmarks (MAE)} \\
    Malonaldehyde & \ce{C3H4O2} & 9 & 6.5 & 307/338-64-64-1 \\
    Ethanol & \ce{C2H6O} & 9 & 6.4 & 151/174-64-64-1 \\
    Uracil & \ce{C4H4N2O2} & 12 & 7.0 & 220/256-64-64-1 \\
    Toluene & \ce{C7H8} & 15 & 6.5 & 774/596-64-64-1 \\
    Salicylic acid & \ce{C7H6O3} & 16 & 5.7 & 655/731-64-64-1 \\
    Naphthalene & \ce{C10H8} & 18 & 6.5 & 784/596-50-100-1 \\
    Aspirin & \ce{C9H8O4} & 21 & 6.2 & 757/732-64-64-1 \\
    \addlinespace[3pt]
    \multicolumn{5}{l}{Reactive benchmarks (RMSE)} \\
    \ce{F + C2H6} & \ce{H6C2F} & 9 & 3.2 & 201/353-30-40-1 \\
    \ce{F^- + (CH3)3CI} & \ce{C4H9FI} & 15 & 15.7 & 500/790-50-100-1 \\
    \ce{Cl^- + (CH3)3CI} & \ce{C4H9ClI} & 15 & 18.1 & 500/790-70-100-1 \\
    syn-\ce{CH3CHOO} & \ce{C2O3H6} & 11 & 27.0 & 685/728-60-65-1 \\
    anti-\ce{CH3CHOO} & \ce{C2O3H6} & 11 & 19.6 & 437/549-50-65-1 \\
    \bottomrule
  \end{tabular}
  \par\smallskip
  {\footnotesize\raggedright Configuration counts are approximate. The values separated by a slash give the FI/RSI input dimensions; the three subsequent values give the widths of the two hidden layers and the single output layer.\par}
\end{table}

\subsection{Data Splitting and Neural-Network Fitting}

For each benchmark, FI and RSI models were fitted using the same 20 fixed training/held-out splits, generated as described in SI Section S10.3. The training fraction was 0.8 for nonreactive systems and 0.9 for reactive systems. The held-out subset was used for early stopping and is referred to as the test set below. The reported accuracy is the minimum error on this subset among the 20 fits, selected independently for each descriptor; the two selected models therefore need not come from the same split. This best-of-20 convention is used consistently in the tables, figures, and accuracy comparisons.

All models use a feedforward neural network with two fully connected hidden layers and the activation function $z/\sqrt{(1+z^2)}$. The hidden-layer widths are matched between FI and RSI for each system, whereas the input dimensions are determined by the respective descriptors (Table~\ref{tab:systems}); the total parameter counts are therefore not matched. Network parameters were optimized with the Levenberg--Marquardt (LM) algorithm using the fitting framework employed by Hao et al.\cite{Hao2025PIPFIAlgorithms} FI and RSI share the same optimization settings and stopping criteria within each benchmark.

\subsection{Accuracy and Timing Measures}

We report the mean absolute error (MAE) for nonreactive systems and the root-mean-square error (RMSE) for reactive systems, both in meV. These measures are analyzed separately. Both salicylic acid fits reached the LM-factor upper limit; this system is excluded from accuracy comparisons but retained in the potential energy evaluation time analysis.

In Figure~\ref{fig:tradeoff}, the change in fitting error is calculated as the best RSI test error minus the best FI test error, so negative values favor RSI.

\begin{sloppypar}
Potential energy evaluation speedup is reported as $t_{\mathrm{FI}}/t_{\mathrm{RSI}}$, where $t$ is the sum of descriptor-construction time and neural-network forward-pass time for all configurations in a complete benchmark dataset. These two stages convert Cartesian coordinates to descriptors and then descriptors to potential energy, respectively. Each stage was timed under the same conditions for FI-NN and RSI-NN within each benchmark, as specified in SI Section S12. Descriptor construction accounts for most of the total time: the median NN contribution is 2.3\% for FI-NN and 9.9\% for RSI-NN across the 12 systems (SI Section S12 and Table~S6). Tables~\ref{tab:nonreactive} and~\ref{tab:reactive} also report the accumulated expression workload $W$, obtained by summing each feature's additive-term count multiplied by its FI-base order (SI Section S10.4). K and M denote $10^3$ and $10^6$, respectively.
\end{sloppypar}

\section{Results}

\subsection{Accuracy--Cost Balance}

\begin{figure}[htbp]
  \centering
  \includegraphics[width=\linewidth]{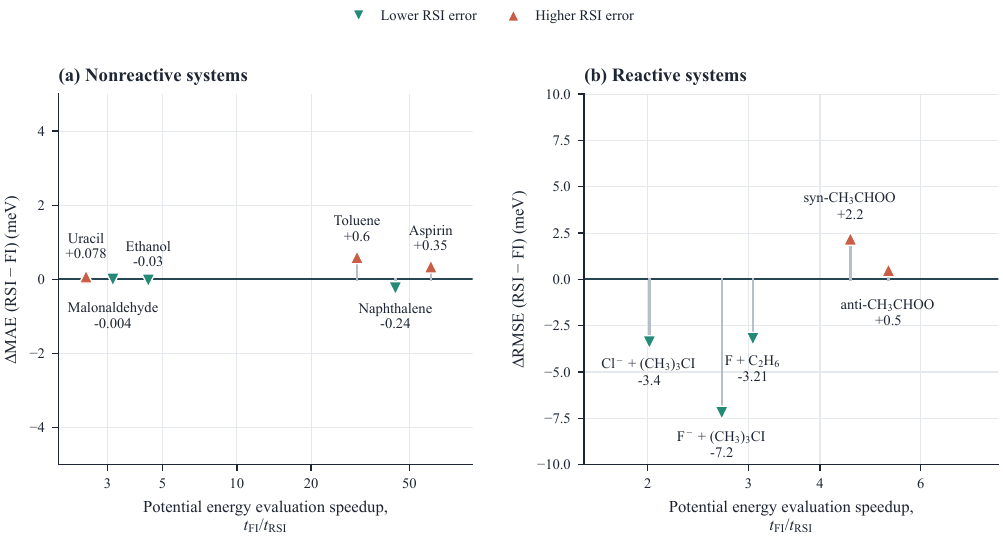}
  \caption{Change in fitting error versus potential energy evaluation speedup for RSI relative to FI. The horizontal axis shows $t_{\mathrm{FI}}/t_{\mathrm{RSI}}$ on a logarithmic scale. The vertical axis shows the difference between the independently selected best test errors from the same 20 fixed splits; labels give the differences in meV. The zero line indicates equal errors. Panel (a) uses nonreactive MAE with an axis range of $\pm5$ meV; panel (b) uses reactive RMSE with a range of $\pm10$ meV. Salicylic acid is excluded because neither fit converged within the LM-factor limit.}
  \label{fig:tradeoff}
\end{figure}

The principal outcome is a reduction in potential energy evaluation cost without a uniform shift in fitting accuracy (Figure~\ref{fig:tradeoff}). Among the 11 benchmarks with successful FI and RSI fits, the best test error is lower with RSI in six cases and higher in five. All 11 show faster potential energy evaluation.

The changes range from \SI{-0.24}{meV} to \SI{+0.60}{meV} in nonreactive MAE and from \SI{-7.2}{meV} to \SI{+2.2}{meV} in reactive RMSE. The significance of these absolute changes depends on the FI error scale, examined below.

\subsection{System Dependence of Fitting Accuracy}

\begin{figure}[htbp]
  \centering
  \includegraphics[width=\linewidth]{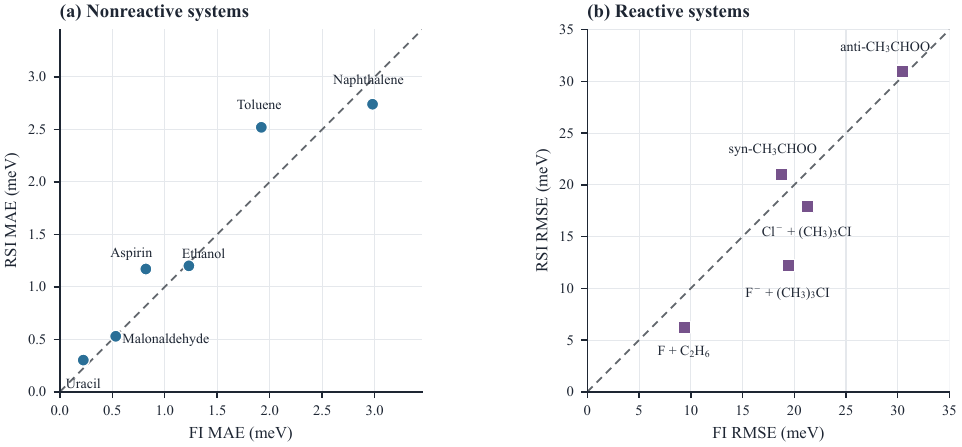}
  \caption{RSI versus FI fitting errors for (a) nonreactive and (b) reactive benchmarks. The dashed diagonal indicates equal errors. Each value is the minimum test error among the same 20 fixed data splits, selected independently for FI and RSI. MAE and RMSE are used for the nonreactive and reactive systems, respectively, in meV. Salicylic acid is omitted because both fits reached the LM-factor upper limit.}
  \label{fig:parity}
\end{figure}

Tables~\ref{tab:nonreactive} and~\ref{tab:reactive} give the numerical results, while Figure~\ref{fig:parity} places the error changes in the context of the FI baseline. In the nonreactive benchmarks, both representations yield errors ranging from sub-meV to a few meV. Their errors are nearly identical for malonaldehyde and ethanol, whereas naphthalene favors RSI. For other systems, the cost reduction involves a discernible accuracy trade-off. For example, the aspirin MAE increases from \SI{0.82}{meV} to \SI{1.17}{meV}: a sub-meV absolute change, but an approximately 43\% relative increase.

\begin{sloppypar}
Among the reactive benchmarks, the lowest test RMSEs obtained with RSI are smaller for all three halogen-containing systems and slightly larger for syn- and anti-\ce{CH3CHOO}. The largest absolute improvement occurs for \ce{F^- + (CH3)3CI}, whose RMSE decreases from \SI{19.4}{meV} to \SI{12.2}{meV}.
\end{sloppypar}

To assess the contribution of bond-angle channels within RSI, we also compared radial-only and radial-plus-angle representations for toluene, aspirin, and \ce{F^- + (CH3)3CI} while keeping the radial features, hidden-layer widths, fitting settings, and 20 data splits unchanged. Under the same best-of-20 convention, adding bond-angle channels reduces the error by 23.4--56.4\% across these three systems (SI Section S11 and Table~S5). This consistent improvement supports the inclusion of angular channels in the benchmark RSI representation.

\begin{table}[htbp]
  \centering\small
  \caption{Accuracy and potential energy evaluation performance for nonreactive benchmarks}
  \label{tab:nonreactive}
  \begin{tabular}{lrrrrrrr}
    \toprule
    System & \multicolumn{3}{c}{FI} & \multicolumn{3}{c}{RSI} & Speedup \\
    \cmidrule(lr){2-4}\cmidrule(lr){5-7}
      & MAE (meV) & $W$ & $t$ (s) & MAE (meV) & $W$ & $t$ (s) & $t_{\mathrm{FI}}/t_{\mathrm{RSI}}$ \\
    \midrule
    Malonaldehyde & 0.533 & 56.5 K & 4.10 & 0.529 & 10.4 K & 1.30 & 3.2 \\
    Ethanol & 1.23 & 56.7 K & 3.32 & 1.20 & 5.7 K & 0.76 & 4.4 \\
    Uracil & 0.224 & 23.7 K & 2.19 & 0.302 & 6.1 K & 0.89 & 2.5 \\
    Toluene & 1.92 & 45.6 M & 4822 & 2.52 & 1.3 M & 157 & 30.7 \\
    Salicylic acid & ERR & 2.6 M & 323 & ERR & 517 K & 28.5 & 11.4 \\
    Aspirin & 0.82 & 11.9 M & 1470 & 1.17 & 282 K & 24.1 & 61.0 \\
    Naphthalene & 2.98 & 158 M & 19399 & 2.74 & 2.9 M & 442.5 & 43.8 \\
    \bottomrule
  \end{tabular}
  \par\smallskip
  {\footnotesize\raggedright ERR denotes a fit that reached the LM-factor upper limit. The definitions of $W$, $t$, and speedup are given in Computational Details.\par}
\end{table}

\begin{table}[htbp]
  \centering\small
  \caption{Accuracy and potential energy evaluation performance for reactive benchmarks}
  \label{tab:reactive}
  \begin{tabular}{lrrrrrrr}
    \toprule
    System & \multicolumn{3}{c}{FI} & \multicolumn{3}{c}{RSI} & Speedup \\
    \cmidrule(lr){2-4}\cmidrule(lr){5-7}
      & RMSE (meV) & $W$ & $t$ (s) & RMSE (meV) & $W$ & $t$ (s) & $t_{\mathrm{FI}}/t_{\mathrm{RSI}}$ \\
    \midrule
    \ce{F + C2H6} & 9.4 & 133 K & 2.963 & 6.19 & 26.6 K & 0.97 & 3.1 \\
    \ce{F^- + (CH3)3CI} & 19.4 & 75 K & 22.8 & 12.2 & 27.7 K & 8.45 & 2.7 \\
    \ce{Cl^- + (CH3)3CI} & 21.3 & 75 K & 17.59 & 17.9 & 27.7 K & 8.73 & 2.0 \\
    syn-\ce{CH3CHOO} & 18.8 & 79 K & 30.5 & 21.0 & 7.7 K & 6.75 & 4.5 \\
    anti-\ce{CH3CHOO} & 30.5 & 169 K & 35.1 & 31.0 & 17.8 K & 6.66 & 5.3 \\
    \bottomrule
  \end{tabular}
\end{table}

\subsection{Potential Energy Evaluation Efficiency}

\begin{figure}[htbp]
  \centering
  \includegraphics[width=0.88\linewidth]{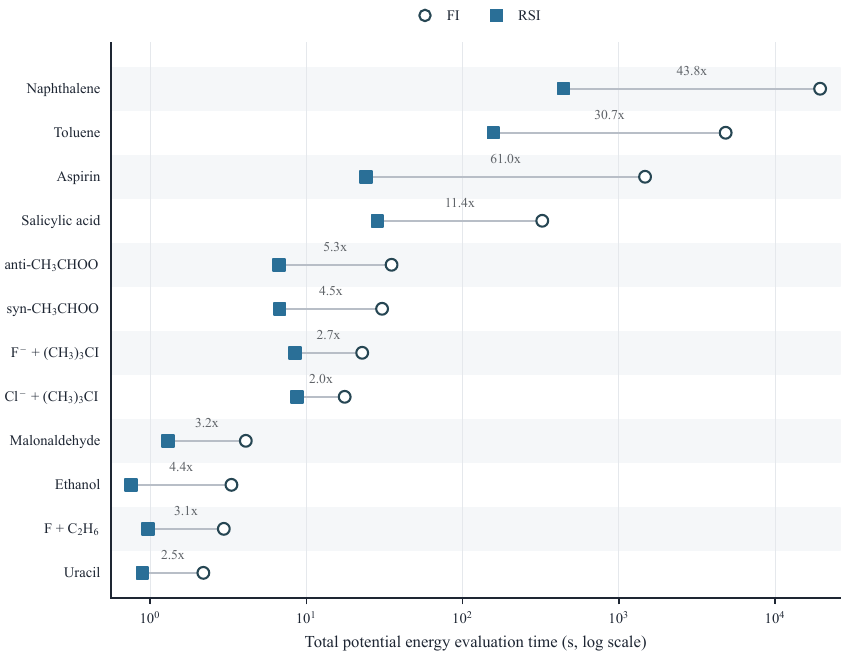}
  \caption{Total potential energy evaluation time for each complete benchmark dataset using FI-NN or RSI-NN. Each total is the sum of Cartesian-to-descriptor conversion time and neural-network forward-pass time. The time axis is logarithmic, and labels give $t_{\mathrm{FI}}/t_{\mathrm{RSI}}$. Each connected pair uses the same configurations and matched timing conditions. Dataset sizes differ between systems; total times are therefore compared within each pair.}
  \label{fig:timing}
\end{figure}

Potential energy evaluation is faster with RSI-NN in all 12 timing comparisons (Figure~\ref{fig:timing}). Speedups span 2.0--61.0, with a median of 4.5. The smaller nonreactive systems and all reactive systems show severalfold improvements, whereas the largest gains occur for nonreactive systems with expensive FI evaluation.

Aspirin has the largest speedup, 61.0, while naphthalene has the largest absolute time saving: potential energy evaluation over its complete dataset decreases from \SI{19399}{s} to \SI{442.5}{s}. These two measures highlight complementary benefits. A severalfold improvement saves relatively little wall time when the initial evaluation takes only seconds; for an expensive FI representation, reducing descriptor-construction cost substantially accelerates the overall potential energy evaluation.

The accumulated expression workload decreases for every system, with a median reduction factor of 7.5. Naphthalene, toluene, and aspirin combine high FI workloads with large timing gains. Workload reduction and measured speedup thus follow the same overall direction, although their magnitudes differ.

\subsection{Configuration-Resolved Error Distributions}

\begin{figure}[p]
  \centering
  \begin{minipage}{0.76\linewidth}
    \raggedright\small
    \textbf{(I) Uracil: RSI-NN}\par\vspace{2pt}
    \includegraphics[width=\linewidth]{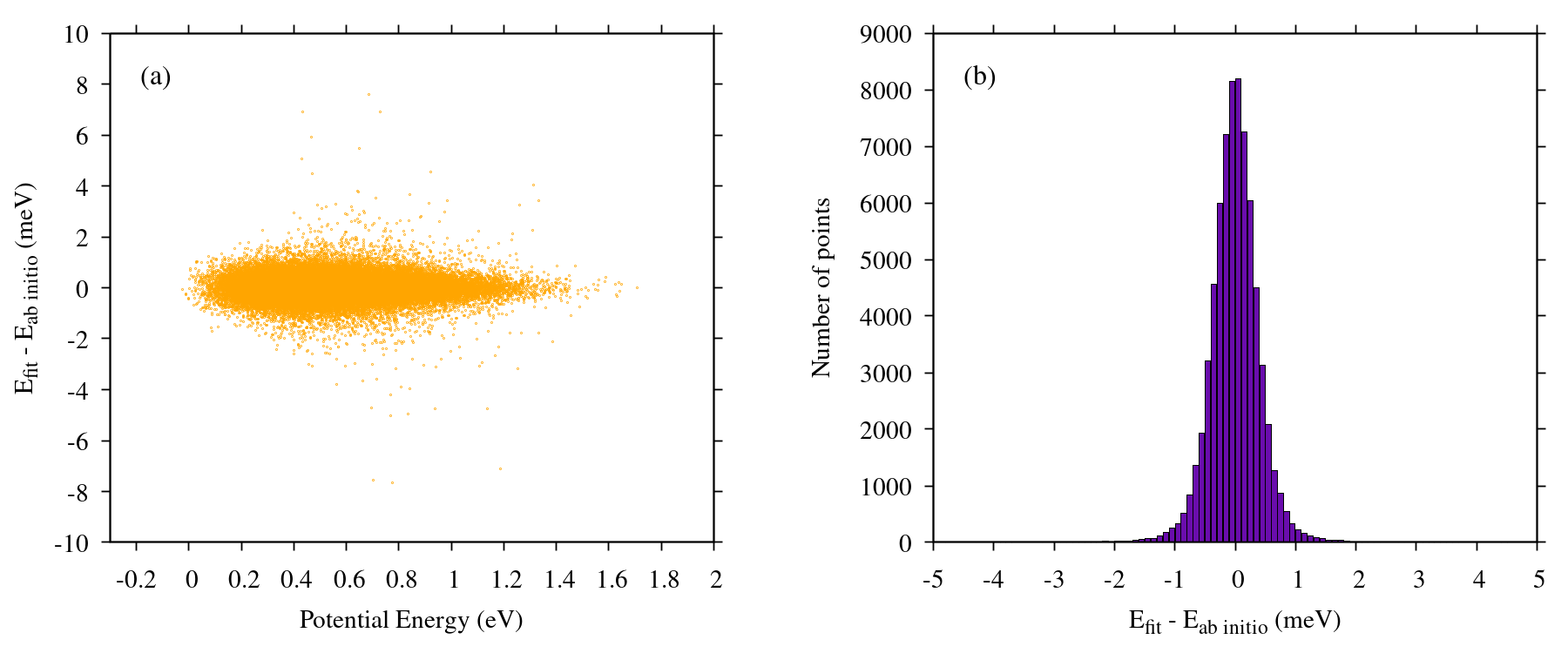}\par\vspace{4pt}
    \textbf{(II) \ce{F^- + (CH3)3CI}: FI-NN}\par\vspace{2pt}
    \includegraphics[width=\linewidth]{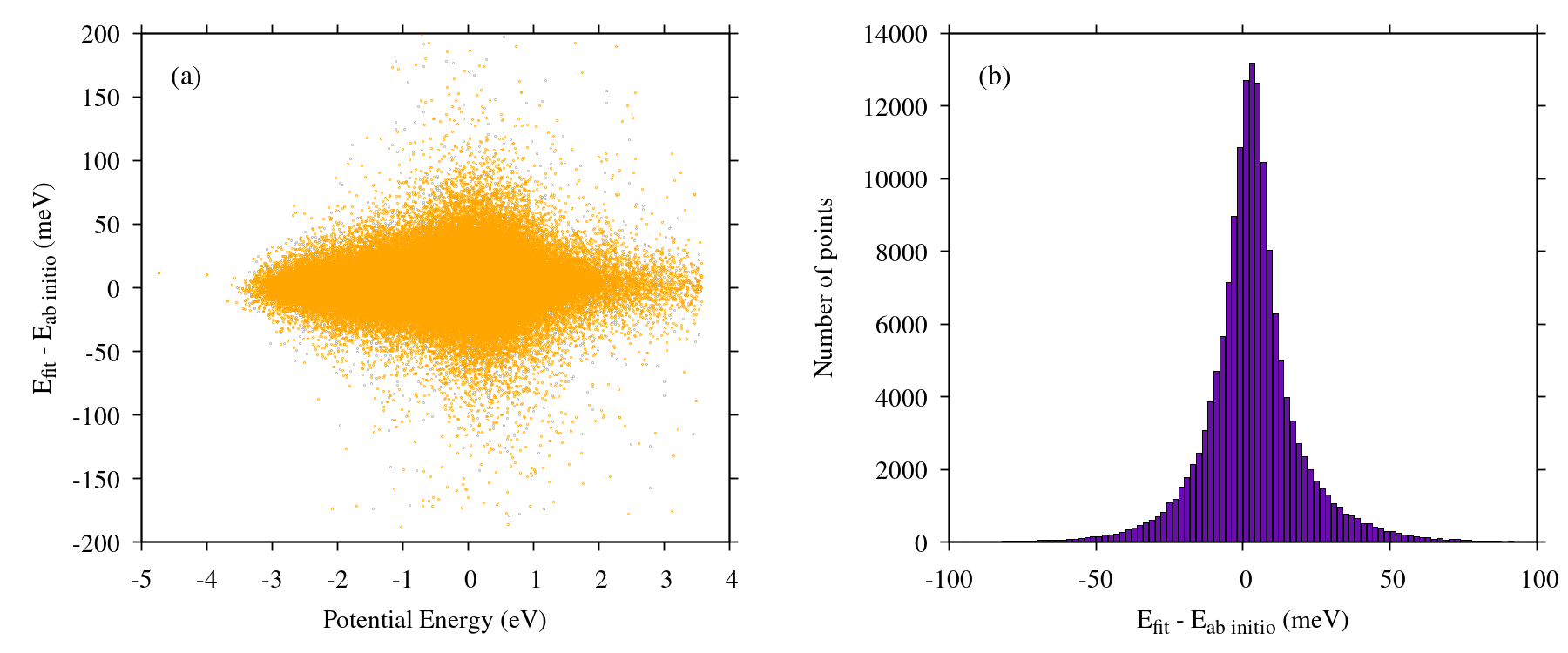}\par\vspace{4pt}
    \textbf{(III) \ce{F^- + (CH3)3CI}: RSI-NN}\par\vspace{2pt}
    \includegraphics[width=\linewidth]{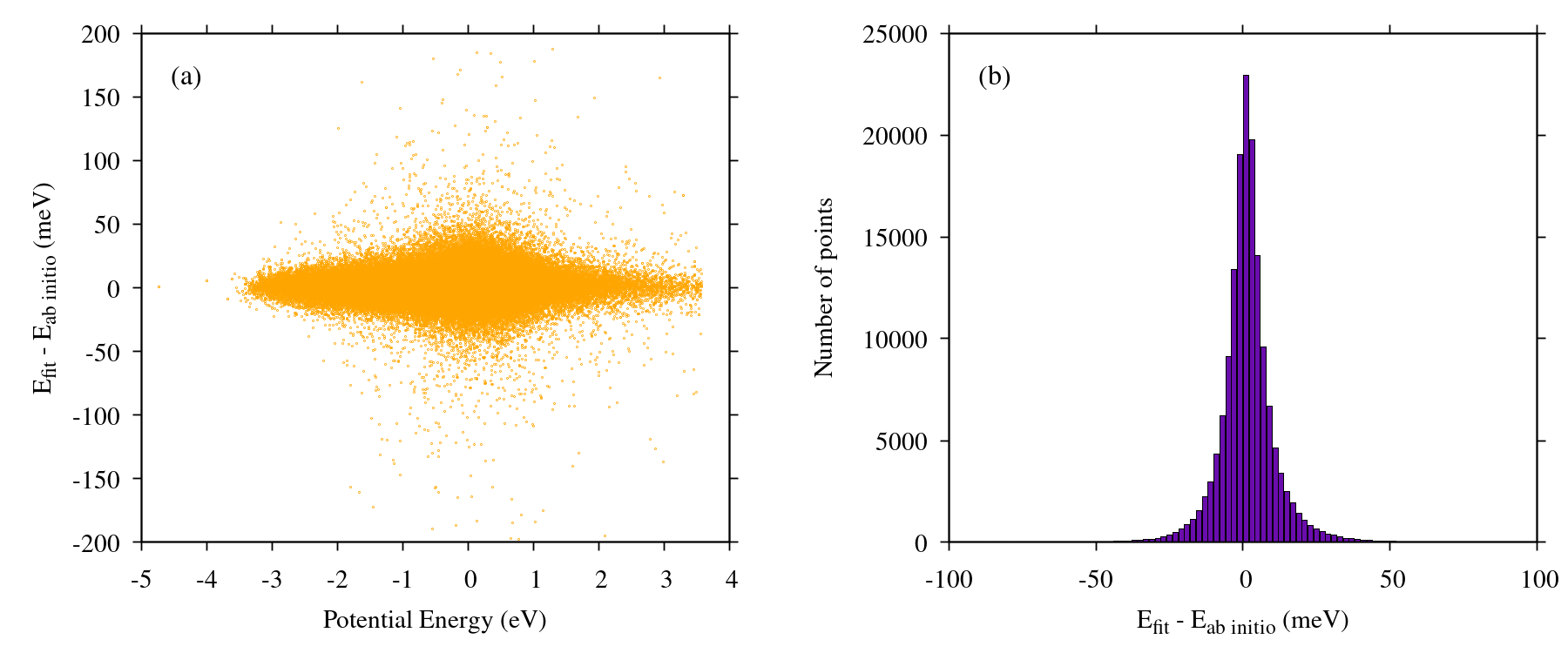}
  \end{minipage}
  \caption{Configuration-resolved residuals over the complete benchmark datasets: (I) RSI for uracil; (II) FI and (III) RSI for \ce{F^- + (CH3)3CI}. Models are selected independently by their minimum test errors. Within each row, (a) shows $\varepsilon=E_{\mathrm{fit}}-E_{\mathrm{ref}}$ versus relative reference potential energy, and (b) shows a residual-count histogram. Potential energies are in eV and residuals in meV. The energy zero is the energy of the first configuration in the chosen split and has no physical significance. The reactive FI/RSI panels share the same reference configuration, scatter-plot limits, and histogram bin width; their histogram count axes are scaled separately. These plots include both training and test configurations.}
  \label{fig:error_distribution}
\end{figure}

Figure~\ref{fig:error_distribution} examines the central distribution and tails underlying the aggregate errors. It shows RSI residuals for uracil and an FI/RSI comparison for \ce{F^- + (CH3)3CI}. For each descriptor, the model with the lowest test error among the 20 fits was evaluated on the complete benchmark dataset, including training and test configurations. Residuals are calculated as fitted energies minus \textit{ab initio} reference energies; positive and negative values indicate overestimation and underestimation, respectively.

For uracil, the RSI residuals form a narrow band around zero, and the histogram has a single dominant peak. A small number of configurations retain residuals of several meV, revealing tails that are not apparent from the MAE alone. For \ce{F^- + (CH3)3CI}, the central scatter band is narrower with RSI than with FI when plotted on the same scales. Histograms with identical bin widths likewise show a more concentrated central distribution for RSI, although substantial positive and negative residuals remain. The lower best test RMSE is therefore accompanied by a narrower central residual distribution over the complete dataset, while the remaining tails identify configurations for more localized assessment.

\section{Discussion}

\subsection{Orbit Separation and Fitting Accuracy}

The theoretical construction and benchmark results address complementary questions. Orbit separation identifies information that an invariant representation should retain, whereas fitting tests assess how effectively a finite descriptor and network use that information. The formaldehyde example shows that algebraic generation can demand more components than orbit separation. The four-atom example then shows why recursive distance power sums alone can miss correlations and why explicit angular information is useful.

The matched radial-only comparisons support this geometric interpretation: adding first-order bond-angle channels improves fitting while leaving the radial representation unchanged. Angular augmentation thus provides an effective way to enrich the retained correlations without expanding the radial basis to higher order. The resulting accuracy relative to FI remains system dependent, reflecting both the selected features and their approximation by the regression network. Together, the angular comparisons and the FI benchmarks support treating correlation content and evaluation cost as linked design choices.

\subsection{Evaluation Cost and Descriptor Size}

The timing gains do not arise from a uniformly smaller input vector. RSI has more input components than FI in nine of the 12 benchmarks (Table~\ref{tab:systems}), yet potential energy evaluation is faster in every case. The dominant saving comes from constructing those components. Recursive scalar functions can be reused in multiple invariant templates, and angular channels can share geometric quantities computed from the same Cartesian coordinates.

The reduced expression workloads are consistent with this organization, especially for naphthalene, toluene, and aspirin. Workload compression alone does not determine wall time: potential energy evaluation also involves geometric setup, channel-specific operations, memory access, descriptor storage, and the neural-network forward pass. The reported speedups account for both descriptor construction and NN evaluation. Descriptor dimension and accumulated expression workload therefore complement, rather than replace, the total potential energy evaluation time.

\subsection{Relation to Existing Methods and PES Applications}

RSI retains the symmetry-adapted-input strategy of PIP-NNs and FI-NNs,\cite{Jiang2013PIPNN,Shao2016FINN} but reorganizes which geometric correlations are retained and how their nonlinear dependence is evaluated. Invariant-generation algorithms expand the range of constructible polynomial sets,\cite{Chen2020FINN2,Hao2025PIPFIAlgorithms} whereas the present recursive construction targets their repeated use across molecular configurations. Its practical aim is to retain useful geometric correlations while limiting the cost of evaluating them.

Expression evaluation is also central to optimized PIP approaches. Houston et al. combined basis compaction and reverse differentiation for potential energy and gradient evaluation and subsequently examined the accuracy--speed balance for aspirin.\cite{Houston2022PIPGradients,Houston2024Aspirin} Here, the timing comparison covers potential energy evaluation from Cartesian coordinates, including descriptor construction and the NN forward pass; force evaluation is outside the present comparison.

The reactive benchmarks connect this computational saving to surfaces used for substitution and elimination reactions, entrance-channel roaming, and rate calculations.\cite{Lu2022Fluoride,Lu2026Chloride,Liu2025Syn,Liu2025Anti} Their system-dependent fitting errors, together with the residual tails, indicate where local assessments near complexes and barriers would be most informative. Gradient timings, force errors, and trajectory tests would establish how the gains in potential energy evaluation translate into dynamical simulations.

\section{Conclusions}

To summarize, we developed recursive separating invariants (RSIs) as an efficient symmetry-preserving descriptor framework for constructing global neural network PESs. Unlike conventional fundamental invariants (FIs), which are designed to generate the complete invariant algebra, RSIs adopt orbit separation as the guiding principle, focusing on retaining the geometric information needed to distinguish symmetry-inequivalent molecular configurations. By combining low-correlation invariant templates with recursive radial functions and complementary bond-angle descriptors, the proposed approach systematically incorporates higher-order nonlinear and geometric information without explicitly constructing large polynomial expansions. This provides a flexible and computationally efficient alternative to conventional FI descriptors while preserving exact permutation symmetry.

The performance of RSI was systematically evaluated on seven nonreactive molecules and five reactive systems, covering molecular systems with up to 21 atoms and reactions involving complex bond rearrangements. Compared with FI-NNs of matched hidden-layer widths, RSI-based models accelerate potential energy evaluation by factors of 2.2–61.8 across all 12 benchmarks, with particularly substantial improvements for larger molecules such as aspirin, naphthalene, and toluene. The fitting accuracy remains system dependent, with RSI yielding lower test errors in six of the eleven successful comparisons and higher errors in five. The overall fitting accuracy is comparable between RSI-NN and FI-NN approaches. These results demonstrate the ability of RSI-NN to substantially accelerate molecular dynamics simulations based on highly accurate global PESs, where repeated evaluations of potential energies and/or forces are required.

\section*{Supporting Information}
Proofs and configuration-reconstruction arguments; explicit four-atom examples; radial and angular recursion and numerical considerations; evaluation-complexity analysis; system-specific descriptor dimensions and recursion settings; data splitting and workload definition; separate descriptor-construction and neural-network timings; matched radial-only and radial-plus-bond-angle comparisons (PDF).

\section*{Acknowledgment}
\begin{sloppypar}
This work was supported by the Strategic Priority Research Program of the Chinese Academy of Sciences (XDB0970203), the National Natural Science Foundation of China (22525304 and 22288201),  the Innovation Program for Quantum Science and Technology (2021ZD0303305), and the Dalian Innovation Support Program (2021RD05).
\end{sloppypar}

\bibliography{ref}

\end{document}

\maketitle

Among the proofs and results below, some are restatements of standard results in invariant theory, and some derive from earlier work; this paper makes no claim of originality for this foundational material. The detailed arguments are listed here mainly to help readers understand the mathematical basis, conditions of applicability, and derivation details of the statements and conclusions in the main text.

\section*{S1. Algebraic completeness implies separability}

This is a standard result in invariant theory; see Kemper\cite{Kemper2009Separating} for the relevant concepts. Let a finite group $G$ act linearly on $\mathbb K^d$ over a field $\mathbb K$ of characteristic zero; the corresponding invariant ring is

\[
\mathbb K[x]^G
=\{h\in\mathbb K[x]:h(gx)=h(x),\ \forall g\in G\}.
\]

If a finite set

\[
F_{\mathrm{gen}}=\{f_1,\ldots,f_L\}
\]

satisfies

\[
\mathbb K[x]^G=\mathbb K[f_1,\ldots,f_L],
\]

then it is said to algebraically generate the invariant ring. If, for any $x,y$ in a given configuration domain $X$,

\[
f_i(x)=f_i(y)\quad(i=1,\ldots,L)
\quad\Longrightarrow\quad
y\in Gx,
\]

then $F_{\mathrm{gen}}$ is said to separate $G$-orbits on $X$.

\textbf{Proposition S1.} For a finite group action, any algebraic generating set is a separating set.

\textbf{Proof.} We first show that invariants of a finite group can separate any two distinct orbits. If $Gx\neq Gy$, then $Gx$ and $Gy$ are two disjoint finite point sets. Since $\mathbb K$ is an infinite field, one can choose a linear function $\ell$ that takes pairwise distinct values on the points of the finite set $Gx\cup Gy$. By univariate Lagrange interpolation, there exists a polynomial $L$ such that

\[
L(\ell(z))=
\begin{cases}
1,&z\in Gx,\\
0,&z\in Gy.
\end{cases}
\]

Let $h=L\circ\ell$, and symmetrize it using the Reynolds operator:

\[
\mathcal R_Gh(z)
=\frac1{|G|}\sum_{g\in G}h(g^{-1}z).
\]

$\mathcal R_Gh$ is a $G$-invariant polynomial that is identically $1$ on $Gx$ and identically $0$ on $Gy$, and can therefore distinguish these two orbits.

If an algebraic generating set $F_{\mathrm{gen}}$ fails to separate some two orbits, then there exist $x\not\sim_Gy$ satisfying $f_i(x)=f_i(y)$. However, any invariant can be written as

\[
h=P(f_1,\ldots,f_L),
\]

so all invariants agree at $x$ and $y$, which contradicts the orbit-separating polynomial constructed above. Therefore

\[
\text{algebraic completeness}\quad\Longrightarrow\quad\text{orbit separability}.
\]

The converse does not hold in general. Orbit separability only requires that the descriptor map be injective on the quotient space $X/G$; it does not require that other invariant polynomials be expressible as polynomials in this descriptor.

\section*{S2. Detailed discussion of the formaldehyde example}

Consider formaldehyde $\mathrm{CH_2O}$, whose nontrivial permutation symmetry is the exchange of the two hydrogen atoms. Write

\[
\begin{aligned}
c&=r_{\mathrm{CO}},&
d&=r_{\mathrm{H_1H_2}},\\
a&=r_{\mathrm{CH_1}}+r_{\mathrm{CH_2}},&
u&=r_{\mathrm{CH_1}}-r_{\mathrm{CH_2}},\\
b&=r_{\mathrm{OH_1}}+r_{\mathrm{OH_2}},&
v&=r_{\mathrm{OH_1}}-r_{\mathrm{OH_2}}.
\end{aligned}
\]

Hydrogen exchange leaves $a,b,c,d$ unchanged, and sends

\[
(u,v)\longmapsto(-u,-v).
\]

Hence the invariant ring in $u,v$ is generated by $u^2,uv,v^2$. Forming the linear combinations

\[
q=u^2-v^2,\qquad
p=2uv,\qquad
s=u^2+v^2,
\]

gives an algebraically complete set of invariants

\[
F_{\mathrm{gen}}
=\{a,b,c,d,q,p,s\}.
\]

These three quantities satisfy

\[
s^2=q^2+p^2.
\]

On the real-variable domain, $s\geq0$, so

\[
s=\sqrt{q^2+p^2}.
\]

Therefore,

\[
F_{\mathrm{sep}}
=\{a,b,c,d,q,p\}
\]

has values that uniquely determine the values of the complete generating set $F_{\mathrm{gen}}$. By the conclusion of Section S1 that an algebraic generating set is necessarily separating, $F_{\mathrm{sep}}$ still separates all real orbits.

Finally, we show that this recovery is indeed not a polynomial relation. Suppose $s=P(a,b,c,d,q,p)$ is a polynomial identity. Since $a,b,c,d,u,v$ are independent linear coordinates, setting $a=b=c=d=0$ gives $s=\widetilde P(q,p)$. Comparing the quadratic homogeneous terms in $u,v$, we must have

\[
u^2+v^2=\alpha(u^2-v^2)+2\beta uv.
\]

Comparing the coefficients of $u^2$ and $v^2$ requires $\alpha=1$ and $-\alpha=1$ respectively, a contradiction. Hence $s\notin\mathbb R[a,b,c,d,q,p]$. Setting these variables to zero here is a check of the identity in the abstract polynomial ring, and does not require that these zero values correspond to valid molecular configurations.

The above discrepancy is not an artifact of choosing a non-minimal generating set. After an invertible linear change of variables, the full invariant ring can be written as

\[
\mathbb R[a,b,c,d]\otimes
\mathbb R[u^2,uv,v^2].
\]

Its space of indecomposable generators contains four first-order generators $a,b,c,d$ and three second-order generators $u^2,uv,v^2$. More precisely, let $A$ be this invariant ring, let $\mathfrak m$ be the maximal ideal of invariants with zero constant term, and let $\mathfrak m^2$ consist of finite sums of pairwise products of such invariants. Since the invariants contain no linear term in $u,v$, these products contain none of the above seven monomials, while every other non-constant invariant monomial can be decomposed into a product of at least two invariants with zero constant term; hence the residue classes of these seven quantities form a basis of $\mathfrak m/\mathfrak m^2$, of dimension $7$. For any algebraic generating set, after subtracting the constant term of each generator, its polynomial combinations leave, in this quotient space, only linear combinations of the generators, so the residue classes of these generators must linearly span the entire seven-dimensional space. Therefore any algebraic generating set contains at least seven elements, and $F_{\mathrm{gen}}$ already attains this lower bound.

The physically realizable, nondegenerate four-atom distances form an open set of dimension six. Quotienting by the finite hydrogen exchange preserves this local dimension. A polynomial descriptor is a continuous map, and a continuous descriptor capable of separating all orbits cannot have fewer than six components. $F_{\mathrm{sep}}$ has exactly six elements, and therefore also attains the minimal number needed for separability on the real configuration domain.

\section*{S3. Power sums recover the unordered multiset of bond values}

Let $x_1,\ldots,x_M$ be $M$ bond-length variables, or an injective transformation thereof, and define the first $M$ power sums

\[
p_n(x)=\sum_{i=1}^{M}x_i^n,
\qquad n=1,\ldots,M.
\]

Let $e_k$ denote the $k$-th elementary symmetric polynomial, with $e_0=1$. Newton's identities read

\[
k e_k
=\sum_{i=1}^{k}(-1)^{i-1}e_{k-i}p_i,
\qquad k=1,\ldots,M.
\]

Hence $p_1,\ldots,p_M$ successively determine $e_1,\ldots,e_M$, and thus determine the monic polynomial

\[
q_x(t)
=t^M-e_1t^{M-1}+\cdots+(-1)^Me_M
=\prod_{i=1}^{M}(t-x_i).
\]

The roots of $q_x$, together with their multiplicities, are exactly the unordered multiset $\{x_1,\ldots,x_M\}$. If two unordered multisets differ, then the corresponding monic polynomials $q_x,q_y$ differ in their roots or multiplicities, so their difference is a nonzero polynomial, and hence there exists some real number $t_0$ such that $q_x(t_0)\neq q_y(t_0)$. If the first $M$ power sums were all equal, Newton's identities would force all coefficients of the two polynomials to be equal, contradicting the value found above; hence there must exist some $n\leq M$ with $p_n(x)\neq p_n(y)$. It follows that

\[
\bigl(p_1(x),\ldots,p_M(x)\bigr)
=\bigl(p_1(y),\ldots,p_M(y)\bigr)
\]

holds if and only if $x$ and $y$ differ only by some permutation of variables in $S_M$. If only $D<M$ power sums are retained, the above complete-recovery conclusion no longer holds in general.

This conclusion can also be derived from algebraic completeness: Newton's identities show that $p_1,\ldots,p_M$ polynomially generate all elementary symmetric polynomials, which in turn generate the ring of symmetric polynomials $\mathbb R[x_1,\ldots,x_M]^{S_M}$; hence the first $M$ power sums are algebraically complete, and by the conclusion of Section S1, they can separate $S_M$-orbits, i.e., they distinguish different unordered multisets.

For multi-element systems, decompose the edge set into first-order variable orbits $E_r=\bigsqcup_\alpha E_\alpha$, and compute in each block

\[
p_{\alpha,n}=\sum_{e\in E_\alpha}x_e^n,
\qquad 1\leq n\leq |E_\alpha|.
\]

This only recovers the multiset within each block, i.e., it separates orbits of the enlarged group $\prod_\alpha S_{E_\alpha}$. Even if two blocks share no edge variables, the same atomic exchange can still act on both blocks simultaneously, so one cannot independently rearrange each block and then assert global configurational equivalence. The CH and OH blocks in formaldehyde are an example of this synchronized action; the mixed quantity $uv$ records pairing information missing from the pure block statistics.

\section*{S4. Generic configuration reconstruction on the domain of physical distances}

The following considers only pairwise distances arising from actual atomic configurations in $\mathbb R^3$, and discusses generic reconstruction of a configuration from its unlabeled distance multiset.

\subsection*{S4.1 Single-element systems}

Here we adopt the algebraic-geometric notion of "general position" or "generic": a property is said to hold generically if it holds outside a proper algebraic set. The reconstruction result of Boutin and Kemper (Theorem 1.6 of ref.~\citenum{Boutin2004Reconstructing}; see also ref.~\citenum{Boutin2007Distances}) shows that, for $N\geq m+2$ points in general position in $\mathbb R^m$, the complete set of unlabeled pairwise distances uniquely determines the point configuration, up to a Euclidean isometry and relabeling of points. Equivalently, there exists a nonzero polynomial $F(Q)$ such that

\[
F(Q)\neq0
\]

implies the configuration can be reconstructed from the distance distribution. Configurations not covered by the theorem are contained in the proper algebraic set $F(Q)=0$. This set has Lebesgue measure zero in $\mathbb R^{mN}$. Hence, as long as the configuration sampling distribution has a continuous density with respect to Lebesgue measure, the probability of landing in the exceptional set is zero.

For a three-dimensional system of identical atoms, $m=3$, so

\[
N\geq5
\quad\Longrightarrow\quad
\begin{gathered}
\text{the complete distance multiset generically}\\
\text{determines the configuration orbit}.
\end{gathered}
\]

If the radial transformation $u(r)$ is injective, then $\{u(r_{ij})\}$ and $\{r_{ij}\}$ carry the same information. Combined with Newton's identities from the previous section, it follows that the first $M$ complete power sums have generic orbit separability for three-dimensional systems of identical atoms with $N\geq5$. The original text uses squared distances; on the domain of nonnegative distances, squaring is injective, and is therefore equivalent to the distance multiset. The exceptional algebraic set is a superset of the actual set of non-reconstructible configurations, not an exact characterization: falling into this set does not necessarily imply non-reconstructibility.

\subsection*{S4.2 Multi-element systems}

The previous section pointed out that recovering bond-length multisets block by block is not, in a general abstract edge space, sufficient to imply global configurational equivalence. Generic reconstruction on the domain of physical distances provides an additional condition: when the total number of atoms is at least $5$, the merged complete distance multiset already generically determines the unlabeled configuration, while the element-wise block structure further retains the categorical assignment of distances.

Suppose the system contains $s$ element types, with $N_\alpha$ atoms of the $\alpha$-th type, and total atom count $N=\sum_\alpha N_\alpha$. The allowed atomic relabeling group is

\[
G=S_{N_1}\times\cdots\times S_{N_s}.
\]

According to the element types of the two endpoints, partition all bond lengths into element-pair blocks $E_{\alpha\beta}$, where $\alpha\leq\beta$. The block sizes are

\[
M_{\alpha\alpha}=\binom{N_\alpha}{2},
\qquad
M_{\alpha\beta}=N_\alpha N_\beta\quad(\alpha<\beta).
\]

For each nonempty block, take the complete power sums

\[
p_{\alpha\beta,n}
=\sum_{e\in E_{\alpha\beta}}u(r_e)^n,
\qquad 1\leq n\leq M_{\alpha\beta}.
\]

Newton's identities separately recover the bond-length multiset within each element-pair block. The total number of power sums required remains

\[
\sum_{\alpha\leq\beta}M_{\alpha\beta}
=\binom N2=M,
\]

unchanged by the number of element types; but the maximum order of each power-sum sequence is reduced from $M$ to the corresponding $M_{\alpha\beta}$, which is typically shorter. More importantly, these block multisets not only contain the merged complete distance multiset, but also retain the element-pair category to which each distance belongs, and thus carry more information than the unlabeled distance data of a single-element system.

When the total number of atoms $N\geq5$, if the block power sums of two multi-element configurations all agree, then their merged complete distance multisets also agree. By the reconstruction theorem of the previous subsection, in the generic case the two are first identical as unlabeled point sets. After further excluding the proper algebraic exception of repeated distances, each distance has a unique correspondence on the reconstructed edges, and its element-pair category is thereby also determined; hence the atomic element labels can only be permuted within atoms of the same element type. Thus the above complete block power sums generically separate orbits of $G$.

Here "more information" refers to retaining the element-pair category assignment, not to imposing an artificial ordering between different bond-length blocks.

\subsection*{S4.3 Scope of applicability}

The above conclusions are subject to the following limitations:

\begin{enumerate}
\item They are generic, not global, conclusions, and cannot rule out strict collisions on special configurations.
\item They do not apply to the three-dimensional case $N=4$; Section S5 gives a physical counterexample.
\item They require complete recovery of the distance multiset, and cannot be directly extended to a finite truncation retaining only $D<M$ low-order moments.
\item Measure zero only describes the probability of a strict collision; it does not guarantee that the inverse mapping is well-conditioned near the exceptional set. Random sampling has nonzero probability of landing in a neighborhood of the exceptional set, and near a collision, different configurations with different energies can have arbitrarily close descriptors, making regression poorly conditioned; at the collision itself, the energy need not be a single-valued function of the descriptor.
\item If sampling is strictly confined to a plane, a fixed bond-length manifold, or a particular reaction path, it is typically singular with respect to the Lebesgue measure on the full Cartesian space. In this case one must examine the intersection of the exceptional set with the actual sampling manifold, and cannot directly invoke the above probability-zero conclusion.
\end{enumerate}

\section*{S5. Explicit verification of the four-atom counterexample and the correlation-order hierarchy}

\subsection*{S5.1 Edge-value assignments and three-dimensional realizability}

Following the numbering in the main text, under the edge order $(12,13,14,23,24,34)$, the three distance vectors are

\[
r_A=(a,a,a,b,b,b),\qquad
r_B=(a,a,b,a,b,b),\qquad
r_C=(a,b,b,a,b,a).
\]

The graphs formed by the $a$ edges are, respectively, a star, a triangle plus an isolated point, and a three-edge path, with degree sequences $(3,1,1,1)$, $(2,2,2,0)$, and $(2,2,1,1)$. When $a\ne b$, this graph must be preserved by any distance-preserving atomic relabeling, so the three configurations are pairwise inequivalent.

Take $a=\sqrt2,\ b=1$. Using the centered Gram matrix criterion, the Cayley--Menger determinant, or direct construction of Cartesian coordinates, one can verify that each of the above three sets of distances corresponds to a nondegenerate three-dimensional tetrahedron. They therefore constitute a valid set of physical counterexamples. Not every pair of positive numbers $a,b$ can simultaneously realize all three configurations; the specific values above already suffice.

\subsection*{S5.2 All second-correlation-order recursions still cannot distinguish the star from the triangle}

Among the distinct edge pairs of four identical atoms, there are only two orbit types: adjacent edge pairs sharing one endpoint, and disjoint edge pairs sharing no endpoint. Table~\ref{tab:si-edge-counts} counts unordered edge pairs, with mixed-valued pairs grouped in the single $(a,b)$ category.

\begin{table}[htbp]
\centering\small
\caption{Edge-pair counts for the four-atom counterexamples}
\label{tab:si-edge-counts}
\begin{tabular}{lrrrrrr}
\toprule
& \multicolumn{3}{c}{Adjacent pairs} & \multicolumn{3}{c}{Disjoint pairs}\\
\cmidrule(lr){2-4}\cmidrule(lr){5-7}
Configuration & $(a,a)$ & $(a,b)$ & $(b,b)$ & $(a,a)$ & $(a,b)$ & $(b,b)$\\
\midrule
$A$: star & 3 & 6 & 3 & 0 & 3 & 0\\
$B$: triangle & 3 & 6 & 3 & 0 & 3 & 0\\
$C$: path & 2 & 8 & 2 & 1 & 1 & 1\\
\bottomrule
\end{tabular}
\end{table}

Each row totals $12+3=\binom62$ pairs. Define

\[
Q_{\mathrm{adj}}=\sum_{\substack{\{e,f\}\\ e,f\text{ share one index}}}x_ex_f,
\qquad
Q_{\mathrm{opp}}=\sum_{\substack{\{e,f\}\\ e,f\text{ share no index}}}x_ex_f.
\]

In this subsection, first take $x_e=r_e$; then

\[
\begin{aligned}
Q_{\mathrm{adj}}(A)=Q_{\mathrm{adj}}(B)&=3a^2+6ab+3b^2,\\
Q_{\mathrm{adj}}(C)&=2a^2+8ab+2b^2,\\
Q_{\mathrm{opp}}(A)=Q_{\mathrm{opp}}(B)&=3ab,\\
Q_{\mathrm{opp}}(C)&=a^2+ab+b^2.
\end{aligned}
\]

So $Q_{\mathrm{adj}}(A)-Q_{\mathrm{adj}}(C)=(a-b)^2\ne0$, meaning the second-correlation-order term can distinguish $C$ from the other two.

If ordered edge pairs are used instead, the counts for same-valued pairs in the table double, and mixed-value pairs split into $(a,b)$ and $(b,a)$; the statistics for $A,B$ remain identical term by term. Hence, for any bivariate function $\psi$, and for any adjacent or disjoint ordered-edge-pair orbit $\mathcal O$,

\[
\sum_{(e,f)\in\mathcal O}\psi(r_e(A),r_f(A))
=\sum_{(e,f)\in\mathcal O}\psi(r_e(B),r_f(B)).
\]

In particular, taking $\psi(s,t)=s^pt^q$, or using any per-bond basis function and its products, cannot break the collision between $A$ and $B$. Any first-correlation-order term is likewise identical, so any number of first- and second-correlation-order orbit sums, together with any deterministic post-processing of them, still cannot separate these two configurations. This is a limitation of fixed correlation order, not of insufficient recursion depth.

\subsection*{S5.3 A single third-correlation-order quantity breaks the remaining collision}

The triangle orbit sum is

\[
T_\triangle(x)=\sum_{1\leq i<j<k\leq4}x_{ij}x_{ik}x_{jk}.
\]

Direct substitution gives

\[
\begin{aligned}
T_\triangle(A)&=b^3+3a^2b,\\
T_\triangle(B)&=a^3+3ab^2,\\
T_\triangle(C)&=2a^2b+2ab^2,
\end{aligned}
\qquad
T_\triangle(A)-T_\triangle(B)=(b-a)^3.
\]

Hence $(Q_{\mathrm{adj}},T_\triangle)$ is already sufficient to distinguish these three examples. If an injective bond-length transformation $x=u(r)$ is used, one need only replace $a,b$ in the above correlation quantities with $u(a),u(b)$; the separation conclusion is unchanged. This only proves that this particular pair of configurations is separated; it does not follow that third-correlation-order terms separate all four-atom configurations.

\subsection*{S5.4 Bond-angle cosines can also carry the missing joint information}

Enumerating all $12$ bond angles, with each atom in turn as center and endpoints unordered, define $A_1=\sum\cos\theta$. For triangles with sides $(a,a,b)$ and $(a,b,b)$, the sums of the cosines of the three interior angles are, respectively,

\[
s_{aab}=1-\frac{b^2}{2a^2}+\frac ba,
\qquad
s_{abb}=1-\frac{a^2}{2b^2}+\frac ab.
\]

When all three sides equal $a$ or all equal $b$, the triangle is equilateral, so

\[
s_{aaa}=s_{bbb}=3\cos\frac{\pi}{3}=\frac32.
\]

The three tetrahedra consist, respectively, of "one $bbb$ face plus three $aab$ faces", "one $aaa$ face plus three $abb$ faces", and "two $aab$ faces plus two $abb$ faces"; hence

\[
\begin{aligned}
A_1(A)&=s_{bbb}+3s_{aab},\\
A_1(B)&=s_{aaa}+3s_{abb},\\
A_1(C)&=2s_{aab}+2s_{abb}.
\end{aligned}
\]

where

\[
\begin{aligned}
A_1(A)-A_1(B)
&=\frac{3(a-b)^3(a+b)}{2a^2b^2},\\
A_1(A)-A_1(C)
&=\frac{(a-b)^2(2a^2-b^2)}{2a^2b^2},\\
A_1(B)-A_1(C)
&=\frac{(a-b)^2(2b^2-a^2)}{2a^2b^2}.
\end{aligned}
\]

Since $a,b>0$, as long as $a\ne b$, the first-order sum of bond-angle cosines alone can directly distinguish $A$ from $B$. However, when $b=\sqrt2a$, $A$ collides with $C$; when $a=\sqrt2b$, $B$ collides with $C$.

Next take the second-order double-angle cosine sum

\[
A_2=\sum_\theta T_2(\cos\theta)=\sum_\theta\cos(2\theta).
\]

Denote the second-order double-angle cosine sums on an $aab$ face and an $abb$ face by $t_{aab}$ and $t_{abb}$, respectively. For a triangle with side lengths $(a,a,b)$, the cosines of the three interior angles are $1-b^2/(2a^2),b/(2a),b/(2a)$. Using $T_2(z)=2z^2-1$, we obtain

\[
\begin{aligned}
t_{aab}
&=2\left[\left(1-\frac{b^2}{2a^2}\right)^2
+2\left(\frac{b}{2a}\right)^2\right]-3\\
&=-1-\frac{b^2}{a^2}+\frac{b^4}{2a^4},\\
t_{abb}
&=-1-\frac{a^2}{b^2}+\frac{a^4}{2b^4}.
\end{aligned}
\]

For the equilateral triangle, $t_{aaa}=t_{bbb}=3\cos(2\pi/3)=-3/2$. According to the face counts of the tetrahedra above,

\[
A_2(A)=t_{bbb}+3t_{aab},\qquad
A_2(B)=t_{aaa}+3t_{abb},\qquad
A_2(C)=2t_{aab}+2t_{abb}.
\]

Hence

\[
\begin{aligned}
A_2(A)-A_2(C)
&=t_{bbb}+t_{aab}-2t_{abb}\\
&=-\frac32+\left(-1-\frac{b^2}{a^2}+\frac{b^4}{2a^4}\right)
-2\left(-1-\frac{a^2}{b^2}+\frac{a^4}{2b^4}\right)\\
&=-\frac{(a^2-b^2)^2(2a^4-b^4)}{2a^4b^4}.
\end{aligned}
\]

Exchanging $a,b$ gives

\[
A_2(B)-A_2(C)
=-\frac{(a^2-b^2)^2(2b^4-a^4)}{2a^4b^4}.
\]

Under the first-order collision condition $b=\sqrt2a$, $A_2(A)-A_2(C)=1/4\ne0$; symmetrically, under $a=\sqrt2b$, we also have $A_2(B)-A_2(C)=1/4\ne0$. Thus the three configurations are distinguished.

This gives an explicit example of recursive enhancement at first correlation order for angular quantities: here, "first correlation order" is with respect to the angle variables themselves, while a single angle already jointly depends on the three distances within a three-atom subconfiguration.

\section*{S6. Fundamental correlation templates and the general recursive construction}

\subsection*{S6.1 Fundamental correlation templates}

For a set $S=\{e_1,\ldots,e_k\}$ consisting of $k$ distinct bond-length variables, define the fundamental correlation template

\[
B_{[S]}(x)=\sum_{S'\in G\cdot S}\prod_{e\in S'}x_e.
\]

Each set of bond-length variables is counted only once in the orbit sum. No monomial contains a second or higher power of the same bond-length variable, so the correlation order and algebraic degree are both equal to $k$. This template is not a synonym for a "minimal algebraic generating set": a fundamental correlation template may also be algebraically redundant.

When variables of the same type use the same basis function, substituting variable by variable directly yields an invariant

\[
B_{[S],n}(Q)=B_{[S]}\bigl(\phi_n(z_1(Q)),\ldots,\phi_n(z_M(Q))\bigr).
\]

If $\phi_n(z)=z^n$, this quantity has exactly algebraic degree $kn$ and correlation order $k$. If $\phi_n$ contains lower-order or constant terms, then upon expansion it may contain lower-correlation terms, and one can only guarantee that the maximum correlation order does not exceed $k$ and the maximum degree does not exceed $kn$.

We refer to the above construction as recursive separating invariants (RSIs). "Recursive" refers to the fact that the univariate basis functions are generated order by order from finite initial values and a recurrence relation in the order $n$; the same intermediate values can be used in multiple invariant templates.

\subsection*{S6.2 Permutation invariance of same-exponent recursion}

Permutation invariance follows from a simple fact: an allowed atomic permutation only relabels the bond-length variables and permutes the monomials within the orbit sum among themselves. As long as all variables that can be exchanged with one another use the same scalar basis and the same recursion order, the set of terms in the sum is unchanged before and after.

Take the second-correlation-order template as an example: let $\mathcal O$ be a set of unordered bond-length pairs that are mutually equivalent under atomic permutation. Same-exponent recursion gives

\[
B_{\mathcal O,p}(x)
=\sum_{\{e,f\}\in\mathcal O}x_e^p x_f^p.
\]

Since the two factors have the same exponent, i.e., $p=q$, exchanging $e$ and $f$ leaves the monomial unchanged; other atomic permutations likewise only rearrange the bond-length pairs within $\mathcal O$. Hence $B_{\mathcal O,p}$ remains permutation invariant.

The same reasoning applies to any correlation order: replacing all same-type variables in a fundamental correlation template uniformly with $\phi_n(x_e)$ only changes the univariate function used for each variable, and does not change the manner of the orbit summation. If two exchangeable factors are given different exponents $p\ne q$, an individual term $x_e^px_f^q$ is in general no longer invariant and would require separate symmetrization; this paper adopts the construction in which same-type variables share a recursion order.

\section*{S7. Radial recursion, degree reduction, and high-order ill-conditioning}

\subsection*{S7.1 Radial variables and function families}

Nondimensionalize each bond length using a fixed scale $\ell>0$ for its element-pair category:

\[
\rho_e=\frac{r_e}{\ell}.
\]

The scale is shared by exchangeable distances and is a manually specified hyperparameter. The construction supports the following four families of radial functions; the benchmark choices are specified in Section S10:

\[
\begin{gathered}
\begin{aligned}
f_n^{\mathrm{poly}}(\rho)&=\rho^n, &
f_n^{\mathrm{rec}}(\rho)&=\rho^{-n},\\
f_n^{\mathrm{exp}}(\rho;\alpha)&=e^{-n\alpha\rho}, &
f_n^{\mathrm G}(\rho;\alpha)&=e^{-n\alpha\rho^2},
\end{aligned}\\
n=1,2,\ldots .
\end{gathered}
\]

Here $\alpha$ is likewise manually specified and is not optimized during model training. For the polynomial and reciprocal functions, $n$ is the ordinary power; for the exponential and Gaussian functions, $n$ is the recursion order of the basis function and simultaneously controls the decay rate. Substituting the same-type radial variables, with the same function and the same recursion order, into the fundamental correlation templates of Section S6 yields descriptors at each order that retain the original permutation invariance and correlation order.

\subsection*{S7.2 Ill-conditioning of high-order power sums}

Take as an example the first-correlation-order power sum of positive variables,

\[
p_n=\sum_{i=1}^{M}x_i^n,
\]

whose high-order ill-conditioning has at least three aspects.

First, if $|x_i|$ deviates significantly from $1$, forming $x_i^n$ directly will grow or shrink exponentially with $n$, causing huge scale differences among different power-sum polynomials as well as floating-point overflow or underflow during internal accumulation.

Second, high-order power sums are increasingly dominated by the extremal variable. Let $a=\max_i x_i>0$, $x_i=a\rho_i$, where $0<\rho_i\leq1$. When the maximum is unique,

\[
p_n=a^n\left(1+\sum_{i\ne i_{\max}}\rho_i^n\right).
\]

The relative contribution of non-maximal terms decays exponentially as $\rho_i^n$, and

\[
p_n^{1/n}\longrightarrow\max_i x_i.
\]

So for large $n$, the value of the polynomial is dominated by the largest $x_i$: if $x_i=r_i$, it is dominated by the largest distance; if $x_i=1/r_i$ or $e^{-r_i}$, it is mainly dominated by the shortest distance.

Third, even in exact arithmetic, although complete power sums can recover the unordered multiset, the inverse recovery may itself be ill-conditioned. The Jacobian of the first $M$ power sums with respect to $x_1,\ldots,x_M$ satisfies

\[
(J_p)_{ni}=n x_i^{n-1},
\qquad
\det J_p=M!\prod_{i<j}(x_j-x_i).
\]

When several bond values are close, the Vandermonde product becomes small, and a small error in the moments may correspond to a large error in the roots; when bond values are exactly equal, this labeled Jacobian becomes rank-deficient. This is a different problem from the dynamic range and extremal dominance encountered in forward evaluation.

\subsection*{S7.3 Root-extraction and logarithmic degree reduction}

For a $d$-th-order descriptor $P_d(Q)$ that can be guaranteed nonnegative, degree reduction by root extraction according to its total order can be performed:

\[
R_d(Q)=P_d(Q)^{1/d}.
\]

For a second-order positive-valued power sum, take the square root; for higher-order power sums, take the corresponding $d$-th root. If $P_d>0$, one may also use logarithmic degree reduction,

\[
L_d(Q)=\frac1d\log P_d(Q)=\log R_d(Q).
\]

On the positive domain, the two are simply invertible, monotonic changes of coordinates:

\[
P_d=R_d^d=e^{dL_d}.
\]

Hence, root-extraction and logarithmic degree reduction do not lose the exact information of any single original descriptor; one can still take the root when $P_d=0$, but the logarithmic form is undefined there.

For a homogeneous polynomial of degree $d$, the scaling behavior of root extraction is clearest:

\[
P_d(cx)=c^dP_d(x)
\quad\Longrightarrow\quad
R_d(cx)=cR_d(x),
\qquad c>0.
\]

It pulls a high-degree polynomial back to a scale approximately linear in the original variables. The logarithmic form instead turns multiplicative scaling into additive shifts; for polynomials corresponding to exponential and Gaussian basis functions, root extraction and the logarithm still compress the dynamic range of the output.

This suggests one potential reason why the two degree-reduction operations may improve fitting: they bring input polynomials of different orders to more similar numerical scales, easing the scale imbalance that the downstream model would otherwise have to handle on its own. Root extraction and the logarithm mainly alleviate the forward numerical-scale problem, and cannot eliminate the extremal dominance or Vandermonde-type inverse problems described in Section S7.2. The benchmark radial features use root-extraction degree reduction, whereas the bond-angle features retain their signed values without degree reduction, as specified in Section S10.

\subsection*{S7.4 Alternative univariate recursions}

The ordinary power basis satisfies

\[
\phi_0(z)=1,\qquad
\phi_1(z)=z,\qquad
\phi_{n+1}(z)=z\phi_n(z).
\]

An alternative approach, not used in the numerical tests of this paper, is to first transform the radial variable within the working interval to $z\in[-1,1]$, and then replace ordinary powers with a bounded polynomial sequence. For example, Chebyshev polynomials of the first kind satisfy

\[
T_0(z)=1,\qquad T_1(z)=z,\qquad
T_{n+1}(z)=2zT_n(z)-T_{n-1}(z),
\]

and Legendre polynomials satisfy

\[
P_0(z)=1,\qquad P_1(z)=z,\qquad
P_{n+1}(z)
=\frac{(2n+1)zP_n(z)-nP_{n-1}(z)}{n+1}.
\]

On $[-1,1]$, both families of polynomials are bounded in value. Since $T_n$ and $P_n$ are both degree-$n$ polynomials with nonzero leading coefficient, when the complete univariate sequence from order zero to order $D$ is retained, they span the same polynomial space as $1,z,\ldots,z^D$; the corresponding change-of-basis matrix is an invertible triangular matrix. Thus, under a complete univariate truncation, the three bases carry the same algebraic information capacity, differing mainly in numerical scale and feature combination.

Neither Chebyshev nor Legendre radial recursion was used in the numerical tests of this paper; they are presented here only as potential alternatives. In actual truncations, the algebraic degree of first-correlation-order polynomials is generally significantly higher than that of the rest, so this kind of alternative recursion is expected to help first-correlation-order polynomials the most; when used in higher-correlation templates, it is also technically more complex than at first order.

\section*{S8. Angular cosines, recursion, and orbit summation}

The benchmark RSI descriptors use bond-angle channels. The dihedral formulas below describe an extension of the construction and are not included in the reported benchmark accuracies or timings.

\subsection*{S8.1 Computing bond-angle cosines}

For three distinct atoms, with $j$ as the center and $\{i,k\}$ an unordered pair of endpoints, let

\[
\mathbf a=\mathbf q_i-\mathbf q_j,\qquad
\mathbf b=\mathbf q_k-\mathbf q_j.
\]

The cosine of the bond angle $\theta\in[0,\pi]$ can be computed directly from the dot product:

\[
z_\theta(j;\{i,k\})
=\frac{\mathbf a\cdot\mathbf b}{\|\mathbf a\|\|\mathbf b\|}
=\frac{r_{ij}^2+r_{jk}^2-r_{ik}^2}{2r_{ij}r_{jk}}.
\]

It can therefore be computed either from Cartesian coordinates or from the three distances via the law of cosines, without first solving for $\theta$. Exchanging the two endpoints leaves its value unchanged, but the roles of center and endpoint cannot be interchanged. Enumerating over all three-atom combinations and choices of center gives a total of $N\binom{N-1}{2}$ bond-angle variables.

\subsection*{S8.2 Dihedral-angle cosines versus signed dihedral angles}

For four distinct atoms, take the central edge $\{j,k\}$ and the endpoint pair $\{i,l\}$, and consider the two planes $(i,j,k)$ and $(l,j,k)$ containing the central edge. Let

\[
\mathbf c=\mathbf q_k-\mathbf q_j,\qquad
\mathbf a=\mathbf q_i-\mathbf q_j,\qquad
\mathbf b=\mathbf q_l-\mathbf q_j,
\]

and define two normal vectors using the same cross-product order:

\[
\mathbf n_i=\mathbf a\times\mathbf c,\qquad
\mathbf n_l=\mathbf b\times\mathbf c.
\]

For the dihedral extension, define the cosine as

\[
z_\phi(\{j,k\};\{i,l\})
=\frac{\mathbf n_i\cdot\mathbf n_l}{\|\mathbf n_i\|\|\mathbf n_l\|}.
\]

This definition holds whenever both normal vectors are nonzero. Exchanging $i,l$ only swaps the two normal vectors; exchanging $j,k$ flips the sign of both normal vectors simultaneously, so neither exchange changes $z_\phi$. Hence the central edge and the endpoint pair can each be treated as an unordered pair, but their roles cannot be interchanged. Enumerating with $j<k,i<l$, the total number of variables is

\[
M_\phi=\binom N2\binom{N-2}{2}=6\binom N4.
\]

The commonly used signed dihedral angle additionally distinguishes the sense of torsion about the central axis, whereas the extension defined above retains only the cosine, so $\phi$ and $-\phi$ give the same variable. "Undirected" here does not mean taking the absolute value of the cosine: when the two endpoints lie in the same half-plane on the same side of the central axis, $z_\phi=1$; when they lie in opposite half-planes, $z_\phi=-1$.

\subsection*{S8.3 Risks of using sine and inverse trigonometric functions}

This paper works directly with cosines and recurses on them, mainly to avoid the derivative singularities that explicit angle or sine representations would introduce at special configurations. For example, recovering $\theta=\arccos z$ from the bond-angle cosine $z$, or constructing $\sin\theta=\sqrt{1-z^2}$, both produce a $\sqrt{1-z^2}$ denominator upon differentiation with respect to $z$, which is singular at $z=\pm1$, corresponding to collinear configurations. Recursing directly on the cosine to obtain higher-order angular functions avoids these intermediate representations entirely, and also avoids having to handle branch choices for signed angles.

The bond-angle cosine itself carries relatively little definitional risk: the usual molecular-configuration domain excludes atomic coincidence, so the two bond lengths in its denominator are nonzero, and hence even if the three atoms are collinear, the cosine and its Cartesian derivatives remain well defined. What is avoided here is the singularity introduced by an additional angle or sine representation, not the exclusion of linear configurations.

The dihedral-angle cosine carries greater risk, because a zero normal vector can arise even without atomic coincidence: if either endpoint is collinear with the two atoms of the central bond, the corresponding plane becomes undefined, which of course also includes the case where all four atoms are collinear. Near collinearity, normalization of the normal vectors can also amplify numerical error and derivatives; using only the cosine and its recursion does not eliminate this geometric degeneracy.

\subsection*{S8.4 Using double-angle cosines instead of powers of cosine in recursion}

For any bond angle or dihedral angle, denote the angle uniformly as $\alpha$, with initial variable $z=\cos\alpha$. If an ordinary power basis is used directly, the $n$-th-order function is $z^n=(\cos\alpha)^n$; the angular construction instead uses the Chebyshev recursion

\[
T_n(z)=\cos(n\alpha)
\]

in place of an ordinary power recursion. The two are not the same function. For example,

\[
T_2(z)=2z^2-1=\cos(2\alpha),\qquad
T_3(z)=4z^3-3z=\cos(3\alpha).
\]

Under an ordinary power recursion, as $n$ increases, the term $z^n$ corresponding to the largest $|z|$ gradually dominates the entire polynomial, while $z^n$ itself tends to zero elsewhere; the resulting polynomial becomes increasingly ill-conditioned as $n$ grows. The Chebyshev recursion instead provides oscillatory angular basis functions at different multiples of the frequency.

From the sum-and-difference formulas for cosine, one obtains the three-term recursion

\[
T_0(z)=1,\qquad T_1(z)=z,\qquad
T_{n+1}(z)=2zT_n(z)-T_{n-1}(z).
\]

Hence, starting from the initial cosine value, all basis functions/monomials $\cos(n\alpha)$ can be generated using only multiplications and additions, without actually computing the angle or repeatedly invoking trigonometric functions at each order. Each order $T_n$ is still a polynomial in the cosine variable, and satisfies $|T_n(z)|\leq1$ on $[-1,1]$.

\subsection*{S8.5 Orbit summation of monomials, as with bond length}

Once the above bond-angle and dihedral-angle cosines are treated as basic variables, their symmetrization principle is exactly the same as for bond length. An atomic permutation $g$ acts on the angle index while preserving the roles of center and endpoint, and satisfies

\[
z_e(gQ)=z_{g^{-1}e}(Q).
\]

For a monomial orbit $\mathcal O$, the orbit sum

\[
I_{\mathcal O}(z)=\sum_{m\in\mathcal O}m(z)
\]

only permutes the terms in the sum under atomic permutation, and is therefore permutation invariant. If the same recursive transformation $z_e\mapsto T_n(z_e)$ is applied to variables of the same type, the resulting $I_{\mathcal O}(T_n(z))$ remains invariant.

The tested representation uses first-correlation-order bond-angle templates; the same construction extends to dihedral variables. The correlation order here is defined with respect to the chosen angle variables, and cannot be directly equated with the correlation order of the distance variables involved in these angles. For the first-order variable orbit $E_{\theta,\beta}$ of bond angles, compute

\[
F_{\theta,\beta,n}(Q)
=\sum_{e\in E_{\theta,\beta}}T_n(z_\theta(e;Q)).
\]

For the first-order variable orbit $E_{\phi,\beta}$ of dihedral angles, compute

\[
F_{\phi,\beta,n}(Q)
=\sum_{e\in E_{\phi,\beta}}T_n(z_\phi(e;Q)).
\]

For a fundamental correlation template containing multiple variable factors, the procedure is likewise to first recurse variable by variable, then multiply and sum according to the original template; the only difference is whether the basic variable is a bond-length function or an angular cosine, not the principle of orbit summation. For multi-element systems, orbits are constructed according to the allowed permutations of atoms of the same element, preserving the type and role of center and endpoint.

\section*{S9. Evaluation complexity and storage under diagonal truncation}

\subsection*{S9.1 Diagonal truncation and scalar-basis partitioning}

Let $K$ be the highest correlation order retained, and let $\mathcal T_k$ be the set of fundamental correlation templates used at order $k$. For each correlation order $k$, multiple univariate scalar bases can be used simultaneously. Denote the categories of these bases by $b\in\mathcal B_k$, let $D_{k,b}$ be the number of scalar basis functions retained in category $b$, and define

\[
D_k=\sum_{b\in\mathcal B_k}D_{k,b}.
\]

$D_k$ is the total number of distinct scalar basis functions used at correlation order $k$, and is also the parameter index of this paper's diagonal truncation at that correlation order. Different radial transformations are not an alternative choice outside the diagonal truncation, but rather a partitioning of the quota within $D_k$. For example, if a given correlation order simultaneously uses reciprocal, exponential, and Gaussian scalar bases, one specifies $D_{k,\mathrm{rec}}$, $D_{k,\mathrm{exp}}$, and $D_{k,\mathrm G}$ separately, and their sum is $D_k$.

Denote all scalar bases of category $b$ used as

\[
\phi^{(b)}_1,\ldots,\phi^{(b)}_{D_{k,b}},
\]

Substituting these functions one by one into the fundamental correlation templates in $\mathcal T_k$ yields the descriptors retained at that correlation order. The core of the diagonal truncation we use is allowing $D_k$ to vary with $k$: lower correlation orders can be allocated more scalar bases, while higher correlation orders use a smaller scalar-basis budget. Thus, the highest correlation order $K$, the total index $D_k$ at each order, and the specific partition of each $D_k$ among different scalar-basis categories jointly determine how the polynomials are used and their total number.

\subsection*{S9.2 General evaluation cost}

The full-distance description has

\[
M=\binom N2=O(N^2)
\]

bond-length variables in total. A $k$-th-order fundamental correlation monomial selects $k$ distinct variables from among these; the monomial orbits of all $k$-th-order fundamental correlation invariants together cover these selections, so the total number of monomials is bounded above, roughly, by

\[
\binom Mk\leq M^k=O(N^{2k}).
\]

Precomputing the $D_k$ scalar basis functions allocated to correlation order $k$ for each bond-length variable requires

\[
O(MD_k)=O(D_kN^2)
\]

recursion operations. If each factor in a $k$-th-order monomial can be independently chosen from these $D_k$ scalar basis functions, then each original monomial corresponds to at most $D_k^k$ combinations. Computing the product of the $k$ factors for each combination and accumulating it into the corresponding invariant, the rough upper bound for direct evaluation is

\[
O\!\left(k\binom MkD_k^k\right)
\subseteq O\!\left(kN^{2k}D_k^k\right).
\]

This estimate is intentionally loose, covering all $k$-th-order templates and the complete set of mixed exponents.

For fixed numbers of basis functions, common upper bounds for direct evaluation are as follows. To simplify notation, the $D$ in the table denotes an upper bound on the scalar-basis budget $D_k$ at the corresponding correlation order; multi-indices are chosen as noted in each case.

\begin{table}[htbp]
\centering\small
\caption{Direct evaluation costs for radial and angular constructions}
\label{tab:si-cost}
\begin{tabular}{p{0.40\linewidth}p{0.29\linewidth}l}
\toprule
Descriptor & Variables or combinations & Cost\\
\midrule
Distance, first correlation order & $M_r=\binom N2$ & $O(N^2D)$\\
Full power sums & $M_r=\binom N2$ & $O(N^4)$\\
Distance, second correlation order, complete mixed exponents & $O(N^4)$ & $O(N^4D^2)$\\
Bond angle, first correlation order & $M_\theta=N\binom{N-1}{2}$ & $O(N^3D)$\\
Undirected dihedral angle, first correlation order & $M_\phi=\binom N2\binom{N-2}{2}$ & $O(N^4D)$\\
\bottomrule
\end{tabular}
\end{table}

\subsection*{S9.3 Streaming recursion and training memory}

The univariate power basis needs to retain only the current order, and the three-term recursion needs only the two adjacent orders. The same-exponent scheme can therefore aggregate immediately at each order, generating all outputs with an $O(L)$ recursive working set, where $L$ is the number of scalar geometric variables; storing the outputs requires an additional $O(n_{\mathrm{feat}})$ space, and the template index must also be stored separately. Complete mixed exponents typically require caching an $O(LD)$ recursion table, or incurring a recomputation cost. Reverse-mode automatic differentiation, which stores the computation graph, may occupy memory related to the total number of operations, and this cannot directly inherit the $O(L)$ conclusion of forward streaming computation.

\section*{S10. Benchmark Descriptor Configurations and Computational Settings}
\label{sec:si-configurations}

This section specifies the descriptor configurations, data-splitting procedure, and evaluation measures used in the benchmark comparisons.

\subsection*{S10.1 Descriptor dimensions and configuration identifiers}

All benchmark RSI representations contain radial and bond-angle channels and exclude dihedral channels. Table~\ref{tab:si-configurations} gives the input dimensions and assigns a system-specific identifier to each configuration. The radial and angular channel counts sum to the listed RSI dimension. The identifiers refer jointly to the system and its retained features, so two systems with the same recursion limits need not have identical feature sets.

\begin{table}[htbp]
\centering\small
\caption{Descriptor dimensions and system-specific configuration identifiers}
\label{tab:si-configurations}
\begin{tabular}{lrrrrl}
\toprule
& \multicolumn{1}{c}{FI} & \multicolumn{3}{c}{RSI} & \\
\cmidrule(lr){2-2}\cmidrule(lr){3-5}
System & Total & Radial & Angle & Total & Config. ID\\
\midrule
Malonaldehyde & 307 & 270 & 68 & 338 & MAL\\
Ethanol & 151 & 126 & 48 & 174 & ETH\\
Uracil & 220 & 180 & 76 & 256 & URA\\
Toluene & 774 & 500 & 96 & 596 & TOL\\
Salicylic acid & 655 & 659 & 72 & 731 & SAL\\
Naphthalene & 784 & 500 & 96 & 596 & NAP\\
Aspirin & 757 & 588 & 144 & 732 & ASP\\
\ce{F + C2H6} & 201 & 257 & 96 & 353 & FET\\
\ce{F^- + (CH3)3CI} & 500 & 486 & 304 & 790 & F-TBI\\
\ce{Cl^- + (CH3)3CI} & 500 & 486 & 304 & 790 & CL-TBI\\
syn-\ce{CH3CHOO} & 685 & 404 & 324 & 728 & SYN\\
anti-\ce{CH3CHOO} & 437 & 425 & 124 & 549 & ANTI\\
\bottomrule
\end{tabular}
\end{table}

\subsection*{S10.2 Radial and bond-angle constructions}

The radial block is generated from bond-length FI bases using the scalar functions $1/r$, $\exp(-r)$, and $\exp(-r^2)$. Their successive recursive powers are $r^{-n}$, $\exp(-nr)$, and $\exp(-nr^2)$, respectively, with $n=1,2,\ldots$.

Table~\ref{tab:si-recursion} lists the maximum recursion orders separately for each base-FI order. Entries separated by vertical bars correspond, from left to right, to first-, second-, and, where present, third-order radial FI bases. A limit $D$ retains recursion orders $1,\ldots,D$ for that family and base-FI order; a zero means that the family is not used at that base order. The number of entries determines the highest retained base-FI order $K$.

For example, the limits \texttt{[6|3, 2|1, 2|1]}, listed in the family order \texttt{[reciprocal, exp, gaussian]}, use first- and second-order radial FI templates. The reciprocal limits are 6 and 3, and the exponential and Gaussian limits are both 2 and 1, respectively. These limits control the scalar functions introduced into the templates; they are distinct from the order of the base FI and from the final descriptor dimension.

\begin{table}[htbp]
\centering\small
\caption{Maximum radial recursion orders and bond-angle harmonic orders}
\label{tab:si-recursion}
\begin{tabular}{lccccc}
\toprule
Config. ID & $K$ & Reciprocal & Exponential & Gaussian & \texttt{cos}\\
\midrule
MAL & 2 & $6\mid3$ & $2\mid1$ & $2\mid1$ & 4\\
ETH & 2 & $6\mid3$ & $2\mid1$ & $2\mid1$ & 4\\
URA & 2 & $4\mid2$ & $2\mid0$ & $2\mid0$ & 2\\
TOL & 3 & $8\mid4\mid2$ & $4\mid2\mid1$ & $4\mid2\mid1$ & 16\\
SAL & 3 & $6\mid3\mid2$ & $2\mid1\mid0$ & $2\mid1\mid0$ & 4\\
NAP & 3 & $8\mid4\mid2$ & $4\mid2\mid1$ & $4\mid2\mid1$ & 16\\
ASP & 2 & $8\mid4$ & $4\mid2$ & $2\mid1$ & 8\\
FET & 3 & $6\mid3\mid2$ & $3\mid1\mid0$ & $3\mid1\mid0$ & 8\\
F-TBI & 2 & $4\mid2$ & $2\mid0$ & $2\mid0$ & 4\\
CL-TBI & 2 & $4\mid2$ & $2\mid0$ & $2\mid0$ & 4\\
SYN & 2 & $6\mid2$ & $2\mid1$ & $2\mid0$ & 4\\
ANTI & 2 & $6\mid3$ & $2\mid1$ & $2\mid1$ & 4\\
\bottomrule
\end{tabular}
\par\smallskip
{\footnotesize\raggedright Configuration identifiers are defined in Table~\ref{tab:si-configurations}. Vertical-bar entries are ordered by the radial base-FI order, not by element-pair category. All systems use only first-order bond-angle FI templates; \texttt{cos} gives their maximum harmonic order.\par}
\end{table}

Function combinations are constructed according to the variable partitions defined by the first-order radial FIs. Within a given base FI, all variables in the same partition share the same scalar-function family and recursion order. Different partitions independently select from the functions allowed at that base-FI order, and all such combinations are enumerated. Each assignment is applied consistently throughout the terms of the base FI, after which the original products and sums are evaluated. This retains permutation invariance while allowing mixed families and recursion orders across distinct partitions.

All systems use only first-order bond-angle FI bases, with the cosine recursion of Section S8.4. Thus, \texttt{cos=4} retains $\cos(n\theta)$ for $n=1,\ldots,4$, equivalently $T_n(\cos\theta)$, within each retained first-order angle template. No higher-order bond-angle FI bases or dihedral channels are included.

FI and radial RSI features undergo root-extraction degree reduction as described in Section S7.3. For FI, the root order is the degree of the invariant in its reciprocal-distance variables. For radial RSI, a positive feature value $P$ is replaced by $P^{1/d}$, where $d$ is the sum of the recursion orders of all scalar factors in a term, including their multiplicities. This effective order generally differs from the order of the base FI. Bond-angle features are not subjected to degree reduction or an absolute-value transformation; both positive and negative values are retained.

The scalar-function families and recursion limits were selected through preliminary fitting tests to obtain a broadly applicable construction. The settings retained for each benchmark are listed in Table~\ref{tab:si-recursion}.

\subsection*{S10.3 Dataset selection and fixed data splits}

The nonreactive configurations were selected using the ordinary Euclidean distance between bond-length vectors. This selection operates on the bond lengths, before their conversion to FI or RSI features.

The FI/RSI comparisons use the same 20 fixed training/held-out splits. The training fraction is 0.8 for nonreactive systems and 0.9 for reactive systems. The random-number generator is initialized once with seed 42 and advances continuously across the 20 splits without reseeding. For a dataset containing $N$ configurations, the index sequence is initialized in the original data order and shuffled as follows.

\begin{verbatim}
sequence = [0, 1, ..., N-1]
for i = N-1, N-2, ..., 2:
    j = random_integer(0, i-1)  # includes 0 and i-1
    swap(sequence[i], sequence[j])

N_train = floor(N * training_fraction)
training = sequence[0 : N_train]
held_out = sequence[N_train : N]
\end{verbatim}

The held-out subset is used for early stopping and is termed the test set in the main-text tables. FI and RSI share the same split indices, and the minimum validation error over the 20 fits is selected independently for each descriptor.

\subsection*{S10.4 Fitting settings, timing, and expression workload}

The network dimensions and fitting framework are given in the main text. FI and RSI use matched hidden-layer widths and the same fitting settings within each benchmark; their total parameter counts differ because their input dimensions differ.

Potential energy evaluation time is the sum of Cartesian-to-descriptor conversion time and the subsequent neural-network forward-pass time for the complete benchmark dataset. FI-NN and RSI-NN were timed under the same conditions for each system. Section S12 and Table~\ref{tab:si-energy-timing} report the two components separately.

The workload contribution of each feature is its number of additive terms multiplied by the order of the FI base from which it is constructed. For RSI, this is the base-FI order, not the effective order after recursion used for degree reduction. In particular, each bond-angle channel uses base order one in this count. The accumulated workload $W$ is the sum of these contributions over all features; K and M in the main-text tables denote $10^3$ and $10^6$. This measure characterizes descriptor expression size, not neural-network workload, and is reported alongside the total potential energy evaluation time.

\section*{S11. Radial-Only and Radial-Plus-Bond-Angle Comparisons}
\label{sec:si-angular-comparisons}

The effect of adding bond-angle channels is assessed for two nonreactive systems, toluene and aspirin, and the reactive \ce{F^- + (CH3)3CI} system.

\subsection*{S11.1 Comparison protocol}

The radial-plus-bond-angle representations are the TOL, ASP, and F-TBI configurations in Tables~\ref{tab:si-configurations} and~\ref{tab:si-recursion}. Each radial-only representation retains exactly the radial block of its corresponding configuration, with 500, 588, and 486 input features, respectively. Adding the angle channels increases these dimensions to 596, 732, and 790.

The radial-only and radial-plus-angle models use identical hidden-layer widths, fitting settings, and the same 20 fixed data splits. For each representation, the minimum validation error among the 20 fits is selected independently. MAE is used for toluene and aspirin, and RMSE for the reactive system, in both cases in meV. The radial-plus-angle values are the same RSI results reported in the main-text benchmark tables.

\subsection*{S11.2 Numerical results}

\begin{table}[htbp]
\centering\small
\caption{Fitting errors with and without bond-angle channels}
\label{tab:si-angular-comparisons}
\begin{tabular}{llrrr}
\toprule
System & Metric & \shortstack{Radial only\\(meV)} &
\shortstack{Radial + angle\\(meV)} & \shortstack{Error reduction\\(\%)}\\
\midrule
Toluene & MAE & 3.29 & 2.52 & 23.4\\
Aspirin & MAE & 1.85 & 1.17 & 36.8\\
\ce{F^- + (CH3)3CI} & RMSE & 28.0 & 12.2 & 56.4\\
\bottomrule
\end{tabular}
\par\smallskip
{\footnotesize\raggedright Each error is the independently selected minimum over the same 20 fixed splits. Percentage reductions are relative to the radial-only error.\par}
\end{table}

Adding bond-angle channels lowers the fitting error in all three systems, with relative reductions of 23.4--56.4\% (Table~\ref{tab:si-angular-comparisons}). The largest improvement occurs for \ce{F^- + (CH3)3CI}, whose RMSE decreases from 28.0 to 12.2~meV. Because the radial features and fitting protocol are unchanged, these comparisons demonstrate the benefit of augmenting the fixed radial representation with angular information. They complement the main-text FI-versus-RSI benchmarks by testing the contribution of bond angles within RSI itself.

\section*{S12. Potential Energy Evaluation Timing Breakdown}
\label{sec:si-energy-timing}

Table~\ref{tab:si-energy-timing} separates the time spent constructing descriptors from Cartesian coordinates and the time spent evaluating the neural-network output from those descriptors. Each component covers all configurations in the corresponding benchmark dataset. The total potential energy evaluation time is their sum; FI-NN and RSI-NN are compared under the same conditions for each system. All timings were measured using a single core of an AMD EPYC 9654 processor.

\begin{table}[htbp]
\centering\small
\caption{Potential energy evaluation times and their two components for complete benchmark datasets}
\label{tab:si-energy-timing}
\begin{tabular}{lrrrrrr}
\toprule
& \multicolumn{3}{c}{FI-NN} & \multicolumn{3}{c}{RSI-NN}\\
\cmidrule(lr){2-4}\cmidrule(lr){5-7}
System & Descriptor (s) & NN (s) & Total (s) & Descriptor (s) & NN (s) & Total (s)\\
\midrule
Malonaldehyde & 3.94 & 0.158 & 4.10 & 1.14 & 0.160 & 1.30 \\
Ethanol & 3.21 & 0.105 & 3.32 & 0.65 & 0.106 & 0.76 \\
Uracil & 2.04 & 0.152 & 2.19 & 0.74 & 0.153 & 0.89 \\
Toluene & 4822 & 0.34 & 4822 & 157 & 0.27 & 157 \\
Salicylic acid & 323 & 0.24 & 323 & 28.2 & 0.26 & 28.5 \\
Naphthalene & 19399 & 0.32 & 19399 & 442.2 & 0.26 & 442.5 \\
Aspirin & 1470 & 0.296 & 1470 & 23.8 & 0.296 & 24.1 \\
\ce{F + C2H6} & 2.911 & 0.052 & 2.963 & 0.905 & 0.065 & 0.97 \\
\ce{F^- + (CH3)3CI} & 22.1 & 0.663 & 22.8 & 7.66 & 0.787 & 8.45 \\
\ce{Cl^- + (CH3)3CI} & 16.75 & 0.837 & 17.59 & 7.75 & 0.984 & 8.73 \\
syn-\ce{CH3CHOO} & 29.3 & 1.22 & 30.5 & 5.48 & 1.27 & 6.75 \\
anti-\ce{CH3CHOO} & 34.5 & 0.60 & 35.1 & 5.96 & 0.70 & 6.66 \\
\bottomrule
\end{tabular}
\par\smallskip
{\footnotesize\raggedright Descriptor denotes Cartesian-to-descriptor conversion; NN denotes the forward pass from descriptors to potential energy. Total times are rounded to the precision used in the main-text tables. Speedups and timing fractions are calculated from the component sums before rounding. These timings do not include force evaluation.\par}
\end{table}

Descriptor construction dominates the total potential energy evaluation time in every benchmark. Across the 12 systems, the median NN contribution is 2.3\% for FI-NN and 9.9\% for RSI-NN. The largest NN fraction is 18.8\%, for RSI-NN in syn-\ce{C2O3H6}. Including the NN forward pass gives potential energy evaluation speedups of 2.0--61.0, with a median of 4.5.

\section*{Additional Mathematical Sources}

The invariant-theory and distance-reconstruction results used above are described in refs.~\citenum{Kemper2009Separating,Boutin2004Reconstructing,Boutin2007Distances}. Recurrence relations and bounds for orthogonal polynomials are available in the NIST Digital Library of Mathematical Functions, \href{https://dlmf.nist.gov/18.9}{Section 18.9} and \href{https://dlmf.nist.gov/18.14}{Section 18.14}.

\bibliography{ref}